\documentclass{svproc}
\usepackage{makeidx}  
\makeindex
\usepackage{url}
\usepackage{amsfonts}

\let\Bbb\mathbb

\begin{document}
\frontmatter          
\pagestyle{headings}  


\title{Scattering Theory in Quantum Mechanical Problems}
\titlerunning{Scattering Theory}  
%
\author{Dmitri Yafaev}
\authorrunning{Dmitri Yafaev} 

\institute{Univ  Rennes, CNRS, IRMAR-UMR 6625, F-35000
    Rennes, France \\
\email{yafaev@univ-rennes1.fr} 
\and
SPGU, Univ. Nab. 7/9, Saint Petersburg, 199034   and    \\ Sirius Math.  Center, Olympiysky av. 1, Sochi, 354349 Russia}

\maketitle              

\begin{abstract}
The aim of the lecture is to briefly describe the mathematical background of scattering theory for two- and three-particle quantum systems.
We discuss basic objects of the theory: wave and scattering operators and the corresponding scattering matrix and illustrate them on the example of the
Schr\"odinger  equation. Our goal is to present  time-dependent and stationary approaches  and to describe the underlying mathematical methods.
We also give a sketch of scattering theory for three interacting quantum particles including a difficult problem of the asymptotic completeness of scattering channels. Along with traditional results, we discuss new  scattering channels arising for long-range pair interactions.

\keywords{wave operators, scattering matrix, scattering channels, asymptotic completeness  }

\end{abstract}

\section{Scattering in two-particle systems}
%
%


 \subsection{Time-dependent Schr\"odinger  equation  and wave operators}

 Mathematical scattering theory is concerned with the
 study of the behavior for large times of solutions of the time-dependent Schr\"odinger  equation 
 \begin{equation}
 i\partial u/\partial t= Hu, \quad u(0)=f \in {\cal H}   ,
\label{eq:1}\end{equation}
 for a quantum system with a self-adjoint Hamiltonian $H$ acting in a Hilbert space $ {\cal H}$.
  Being a part of  the perturbation theory, scattering theory describes  asymptotics of $u(t)$ as $t\to +\infty$ or $t\to-\infty$ in terms of solutions of problem (\ref{eq:1})  with a  ``free"  Hamiltonian $H_{0}$.  Of course, equation
(\ref{eq:1}) has a unique solution $u(t)=\exp(-iHt)f$, and the solution of the same equation
with the   operator
$H_0$ and the initial data $u_0(0)=f_0$ is given by the formula $u_0(t)=\exp(-iH_0t)f_0$.  From the viewpoint of scattering theory
the function $u(t)$ has free  asymptotics as $t\to\pm\infty$ if for an appropriate initial data $f_0^{(\pm)}$ the relation
\begin{equation}
 \lim_{t\rightarrow\pm\infty}\|u(t)-u_0^{(\pm )}(t)\|=0  , \quad    u_0^{(\pm)}(t)=\exp(-iH_0t)f_0^{(\pm)},
\label{eq:2}\end{equation}
holds.
Here and everywhere a relation containing the signs
$``\pm"$ is understood as two independent equalities; $\|\cdot\|$ and $(\cdot, \cdot )$ are the norm and the scalar product  in the Hilbert space $ {\cal H}$. 
 We emphasize that initial data $f_0^{(\pm)}$ are different for $t\rightarrow +\infty$  and $t\rightarrow -\infty$.
 Relation  (\ref{eq:2}) leads to a connection
between the corresponding initial data $f_0^{(\pm)}$ and $f$ given by an equation
\begin{equation}
f= \lim_{t\rightarrow\pm\infty}\exp(iHt)\exp(-iH_0t)f_0^{(\pm)} .
\label{eq:3}\end{equation}

A time dependence of solutions $ u (t)=\exp(-iH t)f $ of problem (\ref{eq:1})  is qualitatively different for eigenvectors $f$ of the operator $H$ and for initial  data $f$ orthogonal to all eigenvectors of $H$ (scattering states).  The subspaces of $\cal H$ spanned by all eigenvectors of the operator $H$ and its orthogonal complement will be denoted ${\cal H}_{p}$ and ${\cal H}_{c}$, respectively, so that
   \begin{equation}
{\cal H}_{p}\oplus{\cal H}_{c}={\cal H}.
\label{eq:dc}\end{equation}
If  $Hf=\lambda f$, then  $u(t)=e^{-i \lambda t}f$. On the contrary, for scattering states $f$, solution     $u(t)$ has necessarily  free asymptotics (\ref{eq:2}) as $t\to\pm\infty$. This fact is known as  the asymptotic completeness.
 
 Mathematical scattering theory requires an advanced classification of the spectrum based on measure
theory.   Let  $E $  be the
spectral family of  a self-adjoint operator $H$.  It turns out that under fairly general assumptions,  the measure $(E(\Lambda)f,f)$ (here $\Lambda\subset{\mathbb R}$ is an interval) is 
absolutely continuous   for all
 $f\in{\cal H}_{c}$, that is, ${\cal H}_{c}$ coincides with the absolutely continuous subspace of the operator $H$.  We denote by $H_{c}$ the restriction of $H$   on  ${\cal H}_{c}$  and by
$P_{c} $   the orthogonal projection on this subspace. 
The same objects for the operator $H_{0}$ will be endowed with the index $``0"$.

Equation (\ref{eq:3}) motivates the following fundamental definition.  Let $H_{0}$ and $H$ be self-adjoint operators.  In applications, the operator $H_{0}$ is usually absolutely continuous, but the operator $H$ may  have  discrete eigenvalues.
{\it  The wave, or M\o ller, operator } $ W^{(\pm)} = W^{(\pm)}  (H,H_0 )$ for the pair  $H_0$  and $H$  is   defined by the relation   
   \begin{equation}
 W^{(\pm)} = \mbox{s-}\!\!\!\lim_{t\to\pm\infty}\exp(iHt)
\exp(-iH_0t)  
\label{eq:8}\end{equation}
provided that the corresponding strong (this means that this relation holds on all vectors $f\in{\cal H}$) limit  
exists.  
   The wave operator is isometric   and  enjoys the intertwining
property
 \[
H W^{(\pm)} =W^{(\pm)}  H_0.  
\]
Therefore   its range ${\rm Ran}\: W^{(\pm)} $ is contained in the absolutely continuous subspace
${\cal H}_{c}$ of the operator $H$.

The operator $W^{(\pm)}  (H,H_0)$ is said to be complete if  the equation   
 \begin{equation}
 {\rm Ran}\: W^{(\pm)}  (H,H_0)={\cal H}_{c} 
\label{eq:10}\end{equation}
holds. 
  It is easy to see that the completeness of $W^{(\pm)}   (H,H_0
)$ is equivalent to the existence of the ``inverse" wave operator $W^{(\pm)}   (H_0,H)$. If the wave operator
 $W^{(\pm)}  (H,H_0)$ exists and is complete, then  the operators $H_c$ and $H_{0}$ are unitarily equivalent.   We
emphasize that scattering theory studies not arbitrary unitary equivalence but only the
``canonical" one realized by the wave operators.

Sometimes it is necessary to introduce a more general definition of wave operators for operators $H_{0}$ and $H$ acting in different spaces
${\cal H}_{0}$ and ${\cal H}$.  Suppose that a bounded operator $J_{\pm}: {\cal H}_{0}\to {\cal H}$  is given. Then   the wave operator  for the triple $H_{0}$, $H$, $J_{\pm}$ is defined by the equality
   \begin{equation}
 W^{(\pm)}  (H,H_{0}; J_{\pm}) = \mbox{s-}\!\!\!\lim_{t\to\pm\infty}\exp(iHt)J_{\pm}
\exp(-iH_0t) .
\label{eq:8J}\end{equation}
Properties of these wave operators are to a large extent similar to those for the particular  case (\ref{eq:8}).

The Schr\"odinger operator $H=-\Delta+V(x)$ in the space
$ {\cal H}=L^2({\mathbb R}^d)$ with a real potential $V$ decaying at infinity is a typical Hamiltonian of scattering theory.
 The spectrum of $H$ consists of the continuous part covering the positive half-line and discrete negative eigenvalues. 
The   operator $H $ describes a particle in an external potential $V$ or two interacting particles.   
Particles may   either form a bound state or be asymptotically free (scattering states). Bound states correspond to eigenvalues of $H$  while scattering states are described by solutions of the stationary Schr\"odinger equation 
  \begin{equation}
 -\Delta \psi  +V(x)\psi=\lambda\psi   
\label{eq:5}\end{equation}
for $\lambda>0$. Scattering theory requires a condition
\begin{equation}
 |V(x)|\leq C(1+|x|)^{-\rho}  
 \label{eq:4}\end{equation}
where at least $\rho>1$. Here and below $C$ are different positive constants whose precise values are inessential.  Let   $H_{0}=-\Delta$. It turns out that under assumption  (\ref{eq:4}) wave operators  (\ref{eq:8}) for the pair $H_{0}$, $H$ exist and are complete. 

The approach discussed briefly  above is called time-dependent. There exists also an alternative, stationary,  approach stated in terms of solutions to stationary   equation (\ref{eq:5}). It is exposed in the next subsection. 


 \subsection {Stationary  scattering theory}

As discussed in the books \cite{LL}, \cite{New1},
   in scattering experiments one sends a beam of
  particles of energy $\lambda>0$   in a direction $\omega$. Such a beam is given by  the   plane wave 
\[
   \psi_0 (x; \omega,\lambda) =\exp(ik\langle \omega,x\rangle),
   \quad k=\sqrt{\lambda}>0,  
      \]
    (which satisfies of
course the free equation $-\Delta\psi_0=\lambda\psi_0$). The scattered particles are described for large distances by  the outgoing spherical    wave 
    \[
    a(\hat{x},\omega;\lambda)
|x|^{-(d-1)/2}
\exp(i k |x |).  
        \]
    Here   $\hat{x}=x |x|^{-1}$ is the direction of observation and {\it the coefficient
    $a(\hat{x},\omega;\lambda)$ is known as the scattering amplitude}. Thus,    quantum particles subject to a potential   $V(x)$ correspond to the solution $\psi$ of stationary Schr\"odinger equation   (\ref{eq:5})
  with asymptotics
\begin{equation}
 \psi(x ; \omega ,\lambda)=\exp(ik\langle \omega,x\rangle)
  +    a(\hat{x},\omega;\lambda)
|x|^{-(d-1)/2}
\exp(ik |x |)+ o(|x|^{-(d-1)/2}) 
\label{eq:6}\end{equation}
 at infinity.
 The existence of such solutions requires of course a proof.
     The differential scattering cross-section defined by the equation
       \[
d \sigma(\hat{x},\omega;\lambda)=   |a(\hat{x},\omega;\lambda)|^2 d \hat{x}   
\]
      gives us the part of particles scattered in a solid angle $d \hat{x}$.

Let 
 \begin{equation} 
        H_0=-\Delta,\quad H =H_0+V(x)  
\label{eq:12}\end{equation} 
be operators in the space ${\cal H} = L^2 ({\mathbb R}^d)$.
     From analytic point of view, the  stationary approach relies on the existence of the  boundary values (in a suitable topology) of the resolvents
      \[R_0(z)=(H_0-z I)^{-1}\quad \mbox{and} \quad
R(z)=(H-zI)^{-1}
\]
 as the spectral
parameter $z$ tends to the continuous spectrum coinciding with $[0, \infty)$. Here and below $I$ is the identity operator. Of course, the limits $R_{0}(\lambda\pm i0) $ and $R(\lambda\pm i0) $ do not exist in the sense of the operator convergence in the space $L^2 ({\mathbb R}^d)$, but this is true for the resolvents  sandwiched by  multiplication operators by sufficiently rapidly decaying functions.  We set
$\langle x \rangle=\sqrt{1+x^2}$  and use the same notation  for  the multiplication operator by this function  of $x$. The key analytic result of stationary scattering theory is that 
 the analytic  operator-valued function
 \begin{equation} 
{\cal R}(z)=\langle x \rangle^{-r} R(z) \langle x \rangle^{-r} 
\label{eq:res3}\end{equation}
for $r>1/2$ is continuous in $z$ up to the cut along the half-axis ${\mathbb R}_{+}$. 
This result is known as {\it the limiting absorption principle}.   One first proves this result for the free resolvent $R_{0} (z)$ and then extends it to $R(z)$ using the resolvent  identity
 \begin{equation} 
 R(z)=R_0(z)- R_0(z) V R (z)=R_0(z)- R (z) V R_0 (z),  \quad \Im z\neq 0,
\label{eq:14}\end{equation}
where $  V=H-H_0$.  

Let us explain this statement for the free resolvent supposing for simplicity that $d=3$.  Then $R_{0} (z)$ is an integral operator acting as
 \begin{equation} 
 \big( R_0(z) f \big) (x)=\frac{1}{4\pi |x-x'|}  \int_{{\mathbb R}^3} e^{i\sqrt{z}|x-x'|} f(x')dx',
 \quad \Im z\neq 0,
\label{eq:res}\end{equation}
where $\Im \sqrt{z} > 0$.  Suppose also that $f(x)$ is a bounded function of compact support.  Then it follows from (\ref{eq:res}) that
 \begin{equation} 
 \big( R_0(\lambda\pm i0) f \big) (x)=\frac{e^{\pm i k |x|}}{4\pi |x|}  \int_{{\mathbb R}^3} e^{\mp ik \langle \hat{x},x'\rangle} f(x')dx' + O \big(\frac{1}{ |x|^2}\big ),
 \quad  k= \sqrt{\lambda}.
\label{eq:res1}\end{equation}
Of course, $R_0(\lambda\pm i0) f \not\in L^2 ({\Bbb R}^3)$, but $\langle x \rangle^{-r}  R_0(\lambda\pm i0) f \in L^2 ({\Bbb R}^3)$ if $r>1/2$. This result remains true for $f$ such that  $\langle x \rangle^{r}   f \in L^2 ({\Bbb R}^d)$ and all dimension $d$ of the space ${\Bbb R}^d$.

To extend this result to operator-valued function (\ref{eq:res3}), one proceeds from the resolvent equation  (\ref{eq:14}) and write it as 
 \begin{equation} 
 {\cal R}(z)= \big(I +  {\cal R}_{0}(z) {\cal V})^{-1}  {\cal R}_{0}(z) 
\label{eq:res2}\end{equation}
where ${\cal R}_{0}(z) = \langle x \rangle^{-r} R_{0}(z) \langle x \rangle^{-r} $, $r=\rho/2$ and $B =  {\cal V} \langle x \rangle^{\rho}$  is  a bounded operator.
As we have already seen,  the operator-valued function $ {\cal R}_{0}(z)$ is continuous up to the positive half-line. Using the self-adjointness of the operator $H$, one can check that the homogeneous equation 
$ f+  {\cal R}_{0}(z)  {\cal V} f=0$ has only a trivial solution $f=0$.  Since the operators ${\cal R}_{0}(z)  {\cal V} $ are compact, the Fredholm alternative implies the existence of the inverse operator in (\ref{eq:res2}). Being put together, the above results on the operator $H_{0}$ yield the  limiting absorption principle for the Schr\"odinger operator $H$.

  In the stationary approach,  one
  changes the time-dependent definition  (\ref{eq:8}) of the
wave operators replacing the unitary groups by the corresponding resolvents $R_0(z)$, 
$R(z)$ and  obtains an expression for $W^{(\pm)}$ in terms of their boundary values. Here the limiting absorption principle plays the crucial role.
In place of   limits  (\ref{eq:8}),   one studies the  boundary values   of the resolvents as the spectral
parameter $z$ approaches  the continuous spectrum.     An important advantage of the stationary
approach is that it gives convenient formulas for the wave operators and the scattering matrix defined in their terms.  

We will first discuss the existence of solutions $ \psi(x ; \omega ,\lambda)$ of equation  (\ref{eq:5}) with asymptotics (\ref{eq:6}).  Suppose  that  condition  (\ref{eq:4}) is satisfied with $\rho> (d+1)/2$.  Using the limiting absorption principle, we can correctly define a solution $ \psi(x ; \omega ,\lambda)$ of equation   (\ref{eq:5}) by the formula
 \begin{equation}
  \psi (  \omega,\lambda) =\psi_0 (  \omega,\lambda)-R(\lambda + i 0) V \psi_0( 
\omega,\lambda).
\label{eq:15}\end{equation}
Resolvent identity  (\ref{eq:14})  implies the Lippmann-Schwinger equation
\[
\psi (  \omega,\lambda)=\psi_0 (  \omega,\lambda)-R_0(\lambda + i 0) V \psi ( 
\omega,\lambda)
\]
which     yields (cf. (\ref{eq:res1})) asymptotics (\ref{eq:6}) with the amplitude
 \begin{equation}
 a(\theta,\omega;\lambda)= \nu_{d}   (\lambda)
 \int_{{\Bbb R}^d} e^{-i k \langle \theta,x\rangle} V(x)\psi (x;
\omega,\lambda) dx . 
  \label{eq:ampl}\end{equation} 
   Here 
   \begin{equation}
 \nu_{d}(\lambda)= -e^{\pi i (d-3)/4}  2^{-1} (2\pi)^{-(d -1)/2}\lambda^{(d-3)/4}
  \label{eq:amplX}\end{equation}
  is a numerical constant; in particular, $\nu_3=- (4\pi)^{-1}$.
     Let us now set
$\xi=\lambda^{1/2}\omega$ ($\xi$ is the momentum variable) and consider two sets of scattering solutions
\[
\psi _{-}(x;\xi)=\psi (x;\omega,\lambda)\quad \mbox{and}
\quad \psi_+ ( x; \xi) = \overline{\psi (x;-  \omega,\lambda)}.
\]
We define
 two transformations
 \begin{equation}
 (\Phi_{\pm} f)(\xi)=(2\pi)^{-d/2}\int_{{\mathbb R}^d} \overline{\psi_\pm (x,\xi)} f(x)dx
\label{eq:17}\end{equation}
  of the space $L^2({\mathbb R}^d)$ into itself. The operators $ \Phi_\pm$ can be regarded  as {\it generalized
Fourier transforms}, and both of them coincide with the usual Fourier transform  $\Phi_0 $ if $V=0$; in this case
$\psi_\pm (x,\xi)=e^{i\langle x,\xi \rangle } $. It follows from equation (\ref{eq:5}) for $\psi (\omega,\lambda)$ that under the action of $\Phi_\pm$ the operator $H$ goes over into the operator $\Omega$ of multiplication by $|\xi|^2$, i.e.,
 \begin{equation}
(\Phi_\pm H f)(\xi) =|\xi|^2 (\Phi_\pm f)(\xi) .
\label{eq:Int}\end{equation}
 Moreover, with the help of equation (\ref{eq:15}), it can be shown that $\Phi_\pm$ is an isometry on ${\cal H}_{c}$, it is
zero on its orthogonal complement $ {\cal H}_{p}$ and its range ${\rm Ran}\: \Phi_\pm =  L^{2}({\mathbb R}^d)$.  This is equivalent to  equations
 \begin{equation}
 \Phi_\pm^\ast \Phi_\pm   =P_{c},\quad  \Phi_\pm    \Phi_\pm ^\ast=I 
\label{eq:18}\end{equation}
where $ \Phi_\pm^\ast$ is the operator adjoint to $ \Phi_\pm$.
Hence any function $f\in {\cal H}^{(c)}$ admits the expansion in the generalized Fourier integral
\[
  f (x)=(2\pi)^{-d/2}\int_{{\mathbb R}^d}  \psi_\pm (x,\xi) (\Phi_\pm  f)(\xi)d\xi.
  \]

The wave operators $ W^{(\pm)} = W^{(\pm)}  (H,H_0) $ for    pair (\ref{eq:12})
admit the
representation
 \begin{equation}
W^{(\pm)}   =   \Phi_\pm ^\ast \Phi_0.
\label{eq:19}\end{equation}
Actually, the existence of limits (\ref{eq:8}) and their properties are consequences of the results on the operators $\Phi_{\pm}$ stated above. Indeed, 
using intertwining property  (\ref{eq:Int}), we find that for an arbitrary $f\in {\cal H}$ and $\hat{f}=\Phi_{0}f$
 \begin{equation}
\| e^{i Ht} e^{- i H_{0}t} f -\Phi_{\pm}^* \Phi_{0}f \|=
\|  e^{- i H_{0}t} \Phi_{0}^*  \hat{f} -\Phi_{\pm}^* e^{- i \Omega t}  \hat{f} \|=
\|  ( \Phi_{0}^*  -\Phi_{\pm}^* )e^{- i \Omega t}  \hat{f}\| .
\label{eq:TD}\end{equation}
Then taking into account  asymptotics (\ref{eq:6}) and integrating by  parts, we check that
  the right-hand side of  (\ref{eq:TD})  tends to zero as $t\to\pm\infty$.
This proves both the existence of the wave operators $ W^{(\pm)}    $  and equality (\ref{eq:19}).
The completeness of $ W^{(\pm)}  $ follows from equation (\ref{eq:19}) and  the first equation (\ref{eq:18}).
The second equality   (\ref{eq:18}) is equivalent to the isometricity of $ W^{(\pm)} $.

 Formula
(\ref{eq:19}) is an example of a stationary  representation for the wave operator.   It formally
implies that
\[
W^{(\pm)} \psi_0(\omega,\lambda)= \psi_\pm(\omega,\lambda),
\]
 which means that each wave operator establishes a one-to-one correspondence
 between eigenfunctions of the continuous spectra of the operators $H_{0}$ and $H$.
 
 
   \subsection{Cook's criterion and smooth method}
   
    Let us now give elementary arguments showing that  wave operators  (\ref{eq:8})  for pairs  (\ref{eq:12}) exist.  Observe that
 \begin{equation}
  \frac{d}{dt}  e^{i Ht} e^{- i H_{0}t} f = ie^{i Ht} Ve^{- i H_{0}t} f, \quad V =H-H_0,
\label{eq:C-}\end{equation}
  and hence limit (\ref{eq:8})  exists if the condition (Cook's criterion)
 \begin{equation}
\int_0^{\pm\infty} \| V \exp(-iH_0t)f\| dt<\infty
\label{eq:C}\end{equation}
is satisfied at least  for vectors $f$ in some set dense in the Hilbert space ${\cal H}  = L^2 ({\mathbb R}^d)$.
 We suppose    $\hat{f}:=\Phi_{0} f\in  C_{0}^\infty ({\mathbb R}^d \setminus\{ 0\})$  so that
  $\hat{f} (\xi)=0$ in some  ball $|\xi |\leq a$.  Let $\chi_{b}$ be the characteristic function of the ball $\{x\in{\mathbb R}^d: |x|\leq b\}$ and note that
  \begin{equation}
\|V \exp(-iH_0t)f\|\leq  \|V \| \|\chi_{a|t|} \exp(-iH_0t)f\|+ \|V  (1 - \chi_{a|t|} )\| \| f\|.
\label{eq:C1}\end{equation}
  Asymptotic behavior of the integral
 \begin{equation}
 \big(\exp(-iH_0 t)f\big) (x) = (2\pi)^{-d/2}\int_{{\mathbb R}^d} e^{ i \langle x, \xi\rangle} e^{-i |\xi |^2 t} \hat{f} (\xi) d\xi 
  \label{eq:C2}\end{equation}
  is determined by  stationary points $\xi$  where  $x- 2\xi t=0$.     Therefore   for $|x| \leq a|t|$, we can integrate $n$ times  by parts in (\ref{eq:C2})  which, for all $n$, yields
    estimates 
   \[
 |\big(\exp(-iH_0t)f\big) (x) | \leq C_{n} (f)(1+|t|)^{-n}
  \]
  with some positive constants $C_{n} (f)$. 
  It follows that the first term on the right in (\ref{eq:C1}) tends to zero as $| t|\to\infty$ faster than any power of $| t|^{-1}$.   The second term on the right  is $O(| t|^{-\rho}) $ due to condition  (\ref{eq:4}). Since $\rho>1$, this implies the convergence of integral (\ref{eq:C}) and hence the existence of the wave operators $W^{\pm}(H,H_{0})$.

We emphasize that this simple result relies on   expression (\ref{eq:C2}) for the evolution operator $\exp(-iH_0t)$.  For the operator $\exp(-i H t)$, such an explicit expression does not exist so that the existence of the wave operators
$ W^{(\pm)}  (H_0,H )$, and hence    of the completeness of  $W^{(\pm)}   (H,H_0)$, require  non-trivial  mathematical tools.
  There are two essentially different  analytic 
approaches in  scattering theory: the trace-class and smooth methods.  

The first of them is basic in the abstract theory of operators. The fundamental result of this method is the Kato-Rosenblum theorem. It states that if the difference $V=H-H_{0}$ of the operators $H_{0}$ and $H$ is trace class (roughly speaking, this means that $V$ is well approximated by operators of finite rank), then the wave operators $W^{(\pm)}   (H, H_0 )$ exist and are complete.

The smooth method that will be used below, we discuss in more details. {\it A bounded   operator $K$  is called $H$-smooth if}
 \begin{equation} 
\int_{-\infty}^\infty \| K \exp(-i H  t)f \|^2 dt \leq C \| f\|^2
\label{eq:13}\end{equation}
{\it for all } $f\in {\cal H}$.    It is important that this definition can be  equivalently reformulated  in terms of the resolvents.  Thus, $K$ is $H$-smooth if and only if 
 \begin{equation} 
 \sup_{\lambda\in{\Bbb R}, \varepsilon >0}
 \| K  \big(R(\lambda+i\varepsilon) -R(\lambda-i\varepsilon)\big)K^*\|<\infty.
\label{eq:16a}\end{equation}

Using definition (\ref{eq:13}), it is easy to show that {\it the wave operators }(\ref{eq:8}) {\it exist and are complete if
\[
V=K K_{0}^*
\]
where the operators $K_{0}$ and $K$ are $H_{0}$- and $H$-smooth, respectively}.  Indeed, it follows  from relation (\ref{eq:C-}) that
 \begin{eqnarray*}
\big| (e^{  i H t_{2}}  e^{- i H_{0}t_{2}} f_{0},f)- (e^{  i H t_1}  e^{- i H_{0}t_1} f_{0},&f&)\big|^2 =\Big| \int_{t_{1}}^{t_{2}} (K_{0}  e^{- i H_{0} t} f_{0}, K  e^{- i H t} f) dt \Big|^2
\\
\leq \int_{t_{1}}^{t_{2}}  &\| & K_{0} e^{- i H_0 t} f_{0}\|^2 dt
\int_{t_{1}}^{t_{2}}  \| K  e^{- i H t} f\|^2 dt  .
 \end{eqnarray*}
 The first integral on the right tends to zero as $t_{1}, t_{2}\to\infty$ because the operator $K_{0}$ is $H_{0}$-smooth. According to  (\ref{eq:13}) the second integral is estimated  by 
 $C\|f\|^2$  which proves the existence of the strong (not only weak) limit in  (\ref{eq:8}). 
Moreover,   under the assumptions above the wave operators $W^{(\pm)} (H,H_{0})$ are unitary.

In applications, the assumption of $H$-smoothness of an operator $K$ imposes too stringent conditions on the operator $H$. In particular, it    excludes its eigenvalues. This drawback can be remedied by the notion of local  $H$-smoothness which looks as a direct generalization of the global notion: an operator $K$ is smooth on an interval $\Lambda\subset {\mathbb R} $ if the operator $K  E(\Lambda)$ is $H$-smooth. A sufficient  condition for $H$-smoothness on  $\Lambda$ is given by estimate (\ref{eq:16a}) where $\lambda$ is restricted on this interval.

Let us specifically consider the pair of operators (\ref{eq:12}) with a potential $V(x)$ satisfying  condition (\ref{eq:4})   for some $\rho>1$.  The limiting absorption principle discussed in Sect.~1.2  shows that the operator $\langle x \rangle^{-r}$ for $r>1/2$ is $H$- and, in particular, $H_{0}$- smooth  on any compact subinterval of ${\mathbb R}_{+}$.
 As explained above, this yields the existence and completeness 
of the wave operators $W^{(\pm)} (H,H_{0})$.

Consistent presentations of the theory exposed in the previous subsections  can be found in the books \cite{RS} and \cite{Y}.

  \subsection{Scattering matrix}

Along with the wave operators,  an important role in scattering theory is played by the scattering operator defined by an equation 
 \begin{equation} 
{\bf S}={\bf S}(H,H_0)= W^{(+)}(H,H_0)^*  W^{(-)}(H,H_0 )  .
\label{eq:11}\end{equation}
 The operator ${\bf S}$ commutes with $H_0$, i.e., ${\bf S}H_{0}=H_{0}{\bf S}$,   and  it is unitary   provided the wave operators $W^{(\pm)}(H,H_0 )$ exist and are complete. The scattering operator
$ {\bf S}(H,H_0) $   connects the asymptotics of the
solutions of equation (\ref{eq:1}) as $t\rightarrow -\infty$ and as $t\rightarrow +\infty$ in
terms of the free problem, that is, $ {\bf S}(H,H_0):f_0^{(-)}\mapsto f_0^{(+)}$ where $f_0^{(\pm)}$ are the same as in equation (\ref{eq:2}). 
Below we consider the pair of operators (\ref{eq:12}) under   assumption (\ref{eq:4}) where $\rho>1$. 
   
   Recall that the operator $H_0=-\Delta$ can    be diagonalized by the classical Fourier transform $\Phi_{0}$: 
    $(\Phi_0 H_{0} f)(\xi) =|\xi|^2 (\Phi_0 f)(\xi) $. The spectral representation of the operator $H_0$ is obtained from the momentum one by  a  change of variables. Actually, let the operator $F_{0}$ be defined by the equality
  \begin{equation}
    (F_{0} f)(\omega; \lambda)= 2^{-1/2} \lambda^{(d-2)/4} (\Phi_{0} f)(\lambda^{1/2}\omega), \quad \lambda>0, \; \omega \in {\mathbb S}^{d-1}. 
  \label{eq:16y}\end{equation}
    The operator $F_{0}$ maps unitarily  the space $L^2 ({\mathbb R}^d)$  onto the space $L^2({\mathbb R}_+;{\mathbb N})$ of vector-valued functions with values in the space $   {\mathbb N}= L^2({\Bbb S}^{d-1})$. The operator $F_{0}$ diagonalizes $H_{0}$, that is, the operator $F_{0} H_{0}F_{0}^*$ acts in the space
    $L^2({\Bbb R}_+;{\mathbb N})$  as multiplication by $\lambda$:
    \[
    (F_{0} H_{0} f) (\lambda)= \lambda (F_{0} f )(\lambda).
    \]
   By definition (\ref{eq:16y}), we have
\[
 (F_0 f)(\lambda)=\Gamma_0(\lambda)f 
\]
   where
    \begin{equation}
  (\Gamma_0 (\lambda) f)(\omega)=  2^{-1/2} \lambda^{(d-2)/4} (2\pi)^{-d/2}\int_{{\mathbb R}^d}\exp \big(-i\lambda^{1/2}\langle x, \omega\rangle\big) f(x)dx.
  \label{eq:16}\end{equation}
  The operator $\Gamma_0(\lambda)$ is well defined, for example, on the Schwartz class and according to the Sobolev trace theorem, for any $r>1/2$, the operator  $\Gamma_0(\lambda)\langle x\rangle^{-r}:  
  L^2({\mathbb R}^d)\to  L^2({\mathbb S}^{d-1} ) $ is bounded and compact. A link of  $\Gamma_0(\lambda)$ with the spectral family of the operator  $H_{0}$ and its resolvent  is given by the formula
   \begin{equation}
  \| \Gamma_0 (\lambda) f \|^2=  \frac{d (E_{0} (\lambda) f,f)}{d\lambda}= \frac{1}{2\pi i}\lim_{\varepsilon\to 0} \big( (R_{0} (\lambda+ i\varepsilon) f, f)
  -(R_{0} (\lambda+ i\varepsilon) f, f)\big).
  \label{eq:16x}\end{equation}

     Since the operator ${\bf S}$ commutes with $H_0$, it
     reduces to multiplication by an  operator-valued function
$S(\lambda)=S(\lambda; H,H_0 ):{\mathbb
N}\to {\mathbb N} $   in the spectral representation of the operator $H_0$:
  \begin{equation}
  (F_0 {\bf S} f)(\lambda)= S ( \lambda )(F_0  f)(\lambda), \quad \lambda>0.
  \label{eq:sm}\end{equation} 
The unitary  operator  $S(\lambda) $  is known as { \it the scattering matrix}. The scattering operator and the scattering matrix are usually of great interest in mathematical
physics problems, because they connect the ``initial" and the ``final" characteristics of the process
directly, bypassing its consideration for finite times. 

Let us obtain a representation
\begin{equation}
 S(\lambda)=I-2\pi i\Gamma_0(\lambda)(V-VR(\lambda+i0)V) \Gamma_0^\ast (\lambda)
\label{eq:33}\end{equation}
for  the scattering matrix in terms of boundary values $R(\lambda+ i0)$ of the resolvent of the operator  $H$.  According to the  limiting absorption principle
the right-hand side of (\ref{eq:33})  is correctly defined as  a bounded operator in the space $ {\mathbb N} $.   It follows from definition (\ref{eq:11})  of the scattering operator  and representation (\ref{eq:19})  for the wave operators   that
 \[
{\mathbf S}= \Phi_0^* \Phi_{+} \Phi_- ^\ast \Phi_0
\]
whence 
 \begin{equation}
 ( \Phi_{+} f)(\lambda)=  S(\lambda)( \Phi_{-} f)(\lambda).
\label{eq:sm2}\end{equation}
Let us write definitions (\ref{eq:15}), (\ref{eq:17}) as
 \[
 (\Phi_{\pm} f)(\sqrt{\lambda}\omega)= (2\pi)^{-d/2}\big(f, (I -R(\lambda\mp i0) V)\psi_{0} (\sqrt{\lambda}\omega)\big),
\]
or using notation (\ref{eq:16}),  as
 \[
 (\Phi_{\pm} f)( \lambda ) = \Gamma_{0}(\lambda) (I - V R(\lambda\pm i0) ) f =:\Gamma_{\pm} (\lambda) f.
\]
Therefore (\ref{eq:sm2}) is equivalent to the equation
 \[
\Gamma_{+} (\lambda) f =  S(\lambda)  \Gamma_{-}(\lambda)  f.
\]
We have to show that this equation is satisfied  by operator (\ref{eq:33}), that is,
 \begin{eqnarray*}
\Gamma_{0}(\lambda) (&I &-V R(\lambda +i0))
 \\
 =  \Big( &I &-2\pi i\Gamma_0(\lambda)(V-VR(\lambda+i0)V) \Gamma_0^\ast (\lambda) \Big)    \Gamma_{0}(\lambda) (I - V R(\lambda -i0) ).
 \end{eqnarray*}
Observe that both sides here contain the common  left factor $\Gamma_{0}(\lambda)$ which can be neglected. Using identity (\ref{eq:16x}) for $\Gamma_0(\lambda)^\ast      \Gamma_{0}(\lambda)$, we see that it suffices to check the relation
 \begin{equation}
  R(\lambda +i0)  - R(\lambda -i0) =   \big(I-R(\lambda+i0)V\big)  \big(  R_{0}(\lambda +i0)  - R_{0}(\lambda -i0)\big)  \big( I-VR(\lambda-i0)\big) .
\label{eq:sm7}\end{equation}
In view of the limiting absorption principle we can replace here $\lambda\pm i0$ by $\lambda\pm i \varepsilon$ for $\varepsilon>0$. Then (\ref{eq:sm7}) is a direct consequence of the resolvent identity (\ref{eq:14}) and of the Hilbert identity 
\[
  R(\lambda +i\varepsilon)  - R(\lambda -i\varepsilon) =2i\varepsilon   R(\lambda +i\varepsilon)   R(\lambda -i\varepsilon).
  \]
  This concludes the proof of representation (\ref{eq:33}).

Putting together representations  (\ref{eq:ampl})  and (\ref{eq:33}), we see that $S(\lambda)-I$ {\it is the
integral operator whose kernel is the scattering amplitude}: 
\[
  (S(\lambda)f)(\theta)=f(\theta) +2 i \lambda^{1/2} \overline{\nu_{d} (\lambda) }  
\int_{{\Bbb S}^{d-1}}a(\theta,\omega;\lambda) f(\omega) d\omega
\]
where the numerical factor $\nu_{d} (\lambda) $  is given by (\ref{eq:amplX}).
Note that the operator $S(\lambda)-I$ is compact.

 
 It follows that  the spectrum of the operator $S(\lambda)$ consists of eigenvalues of finite multiplicity, except
possibly the point $1$, lying on the unit circle and accumulating to  the point $1$ only.
Eigenvalues of $S(\lambda)$ play the role of scattering phases or shifts considered often  for radial potentials $V(x)=V(|x|)$.

 The scattering amplitude    is singular on the diagonal $\theta=\omega$ only. This singularity is stronger for potentials with slower decay at infinity (for $\rho$ smaller). If  $\rho>( d+1)/2$, then   the operator $S(\lambda)-I$ belongs to the Hilbert-Schmidt class. In this case the total scattering cross section    
 \[
\sigma( \omega;\lambda)= \int_{{\Bbb S}^{d-1}}|a(\theta,\omega;\lambda)|^2 d \theta
 \]
 is finite  for all energies $\lambda>0$ and all incident directions $\omega\in {\Bbb S}^{d-1}$. If  $\rho> d$, then   the operator $S(\lambda)-I$ belongs to the trace class. In this case, the scattering amplitude $a(\theta,\omega;\lambda)$ is a continuous function of $\theta,\omega\in{\Bbb S}^{d-1}$
(and $\lambda>0$). The unitarity of   the operator $S(\lambda) $ implies the optical theorem
 \[
 \sigma( \omega;\lambda)= \lambda^{-1/2} \Im \big( \nu_{d}^{-1}(\lambda)a(\omega,\omega ;  \lambda)\big)  .
 \]

Using   resolvent identity (\ref{eq:14}), one deduces from equation (\ref{eq:33}) the Born expansion
\begin{equation}
 S(\lambda)=I-2\pi i\sum_{n=0}^\infty (-1)^n \Gamma_0(\lambda)V(R_0(\lambda+i0)V)^n \Gamma_0^\ast
(\lambda).
\label{eq:33A}\end{equation}
This series is norm-convergent  for small potentials $V$ and
  for high energies $\lambda$.
  
 A sufficiently detailed presentation of  mathematical theory of the scattering matrix is given in Chapter~8 of  the book  \cite{Y}.

  \subsection{ Long-range interactions}

Potentials $V(x)$ decaying   at infinity    as   the
 Coulomb potential 
 \[
 V(x)=\kappa |x|^{-1},\quad d\geq 3,
 \]
or slower  are called long-range. More precisely, it is required that
\begin{equation}
|\partial^\alpha V(x)|\leq
C_{\alpha}(1+| x|)^{-\rho-|\alpha|}, \quad \rho\in (0,1],
\label{eq:LR1}\end{equation}
 for all derivatives $\partial^\alpha V(x)$ of $V$ up to some order. In the
long-range case the wave operators
$W^{(\pm)} (H,H_0)$ do not exist, and  the asymptotic dynamics $\exp(-iH_0 t)$  in definition (\ref{eq:8})  should be
properly  modified.  It can be
done   in a time-dependent way either in the coordinate or momentum
representations. For example, in
the coordinate  representation   the free evolution
$\exp(-iH_0 t)$ should be replaced  by
unitary operators
$U_0(t)$ defined by
 \begin{equation}
 (U_0 (t)f )(x) = \exp (i\Xi (x,t)) (2it)^{-d/2} \hat{f } (x/(2t)),
\label{eq:8M1}\end{equation}
 where $\hat{f }$ is the Fourier transform of $f$. For short-range
potentials we can set
$\Xi (x,t)=(4t)^{-1}|x|^2$.
 In the long-range case   the phase function $\Xi (x,t)$
 is  chosen as  a  (perhaps, approximate) solution of the eikonal
equation
 \begin{equation}
\partial \Xi/\partial t+ |\nabla \Xi|^2+ V=0.
\label{eq:Eik}\end{equation}
In particular, we can set 
 \begin{equation}
\Xi (x,t)=(4t)^{-1}|x|^2-t\int_{0}^1 V(sx)ds
\label{eq:8M2}\end{equation}
if $\rho> 1/2$ in condition (\ref{eq:LR1}). For the Coulomb potential,   we  have
\[
\Xi (x,t)=(4t)^{-1}|x|^2- \kappa t |x|^{-1} \ln |t|
\]
(the singularity at $x=0$ is inessential here). According to  (\ref{eq:8M1}) both in short- and long-range cases solutions $\big( U_{0} (t) f\big) (x)$ of the time-dependent
Schr\"odinger equation  
``live" in a region of the configuration space where $|x|$ is of order
$|t|$. Long-range
potentials change only asymptotic phases $\exp (i\Xi (x,t)) $ of these solutions.  Using the Cook criterion, one can easily prove that, with definition
(\ref{eq:8M1}), {\it  the modified wave operators}
 \begin{equation}
 W^{(\pm)} = \mbox{s-}\!\!\!\lim_{t\to\pm\infty}\exp(iHt) U_{0} (t)
\label{eq:8M}\end{equation}
exist.
As in the short-range case, the operators $  W^{(\pm)} $ are isometric and enjoy the intertwining property (\ref{eq:5}).
Asymptotic completeness (\ref{eq:10})  also remains true. In view of equality (\ref{eq:8M1})  this means that, for all $f\in{\cal H}_{c}$,
 \begin{equation}
 ( \exp(-iHt)  f )(x) = \exp (i\Xi (x,t)) (2it)^{-d/2} \hat{f } (x/(2t))+ \varepsilon (x,t),
\label{eq:8M3}\end{equation}
where $ \varepsilon (x,t)\to 0$ as $t\pm \infty$ in the space $L^2 ({\Bbb R}^d)$. Thus, both for short- and long-range potentials, solutions $ ( \exp(-iHt)  f )(x)$ of the time-dependent Schr\"odinger equations ``live" in the region where $|x|\sim |t|$, that is,  $|x|$ and $ |t|$ are of the same order. Only the phase factors in (\ref{eq:8M3}) are changed.

For long-range potentials, a  proof of the asymptotic completeness is a substantial mathematical problem.
Below we briefly describe one of its solutions exposed more thoroughly in Chapter~4 of \cite{LNM}.  The crucial difference with the short-range case is that now the Schr\"odinger operator $H$ cannot be treated as a perturbation of the kinetic energy  operator $H_{0}$. In particular, this means that although the limiting absorption principle (see Sect.~1.2) remains true, its proof cannot be deduced from that for the operator $H_{0}$.

It now relies on a commutator method. 
Let us illustrate this  method  by the following elementary assertion.
Suppose that
\begin{equation}
\| G f\|^2\leq i([H,B ]f,f) +\| K f\|^2,\quad f\in {\cal H},
\label{eq:SM2}\end{equation}
 where   $B$ is   a bounded operator, the commutator $[H,B]:=H B-  BH$ and $K$ is
$H$-smooth. Then
$G$ is also $H$-smooth. 
 Indeed, it follows from (\ref{eq:SM2}) that for all $f $ and any $T>0$
\begin{equation}
\int_{-T}^T \|G e^{-i Ht}f \|^2 dt \leq i\int_{-T}^T ([H,B] e^{-i Ht}f,e^{-i Ht}f) dt+\int_{-T}^T  \|K
e^{-i Ht}f \|^2 dt.
\label{eq:sm2j}\end{equation}
The first term in the right-hand side equals
\[
\int_{-T}^T d( B e^{-i Ht}f,e^{-i Ht}f)/dt\; dt=( B e^{-i Ht}f,e^{-i Ht}f)\Big|^T_{-T},
\] 
and hence is bounded by $2 \| B \|\: \|f \|^2$. In view of the $H$-smoothness of the operator $K$,  the second term on the right in (\ref{eq:sm2j}) is bounded by $C \|f \|^2$, so that the
left-hand side of (\ref{eq:sm2j}) admits the same estimate.  This result is of interest already in the case $K=0$. Moreover, it allows one to find new
$H$-smooth operators $G$ given an  $H$-smooth operator $K$. 


In applications to  Schr\"odinger operators, one has to consider
the commutator   of   $H$ with
the generator of dilations
\[
{ A}=-i\sum_{j=1}^d (x_{j}\partial_{j}+\partial_{j}x_{j})
\]
which is an unbounded operator.
In view of the relation
\begin{equation}
i[H_{0}, { A}]=4H_{0},
\label{eq:CR}\end{equation}
the operator $ A $ is often called conjugate to the operator $H_{0}$. To a large extent, this relation plays the same role as the commutation relation for the operators of differentiation and  multiplication by an independent variable. A generalization of (\ref{eq:CR}) to the Schr\"odinger operator $H$ is given by an estimate (the Mourre estimate):
\begin{equation}
  c(\lambda) E(\Lambda_\lambda) \leq iE(\Lambda_\lambda)[H,  {  A} ]E(\Lambda_\lambda) ,
\label{eq:mourre}\end{equation}
where $\lambda>0$ is  arbitrary,   $\Lambda_\lambda=(\lambda-\varepsilon,\lambda+\varepsilon)$ and positive numbers $\varepsilon$, $c(\lambda) $ are small enough. Roughly speaking, estimate (\ref{eq:mourre}) means that the observable
\[
(Ae^{-iHt } f, e^{-iHt } f)
\]
is a strictly increasing function of $t$ for all $f\in{\cal H}_{c}$.
An important difference between  (\ref{eq:SM2})  and  (\ref{eq:mourre}) is that the operator  ${  A}$ is unbounded.  This considerably complicates the proof of the limiting absorption principle stating that the operator $\langle x \rangle^{-r}$ is $H$-smooth if   $r>1/2$. 

This result would have implied the asymptotic completeness if it were true for $r =1/2$. Unfortunately, as show formulas (\ref{eq:res1}) or (\ref{eq:8M1}), this is not the  case even for the free operator $H_{0}$. To  fix the problem, we supplement the limiting absorption principle  with a so-called radiation estimate. To
state it, denote by
\[
(\nabla^\bot u)(x)= (\nabla  u)(x) -   |x|^{-2}   \langle (\nabla  u)(x), x \rangle x
\]
   the orthogonal projection in ${\mathbb R}^d$ of the gradient $(\nabla  u)(x)$
on the plane orthogonal
to $x$.  We  show that the operator 
$ G = \langle x\rangle^{-1/2} \nabla^\bot $
 is  $H$-smooth  on any compact $\Lambda\subset (0,\infty)$.  This result  is
formulated as an estimate
(either on the resolvent or on the unitary group of $H$)    which we refer
to as the radiation
estimate. This estimate is not very astonishing from the viewpoint of
analogy with the classical
mechanics. Indeed, in the case of free motion,  the vector
$x(t)$ of the position of a particle  is directed asymptotically as its
momentum $\xi$.
Regarded as a pseudo-differential operator,
$\nabla^{\bot}$ has   symbol
$\xi-|x|^{-2}   \langle   \xi,x \rangle x$
 which equals   zero if $x=\gamma\xi$ for some
$\gamma\in{\Bbb R}$.
Thus,  $\nabla^{\bot}$ removes the part of the phase space where a classical
particle propagates.  Formula (\ref{eq:res1})  shows that the operator $\nabla^\bot $ improves the fall-off the function $ \big( R_0(\lambda\pm i0) f \big) (x)$ as $|x|\to\infty$: $\big(  \nabla^{\bot}  R_0(\lambda\pm i0) f \big) (x)=O (|x|^{-2})$  while  $ \big( R_0(\lambda\pm i0) f \big) (x)$ decays as $ |x|^{-1}$  only. 

To be precise, for a proof of the radiation estimate,   we   use inequality  (\ref{eq:SM2})  with
a differential operator 
\begin{equation}
B=-i \sum (b_{j}\partial_j + \partial_j  b_{j}) , \quad
b_{j}=\partial b/\partial
x_j.
\label{eq:comm1}\end{equation}
We set $b (x)=|x|$ so that the coefficients $b_{j}$ of the operator $B$ are bounded.  The commutator of $H_{0}$ with $B$ is quite explicit,
\[
i[H_{0},B]= 4\sum_{j,k} D_{j}  b_{j,k} D_{k}-(\Delta^2 b), \quad
b_{j,k}=\partial^2 b/\partial x_{j} \partial x_k,
\]
whence
\[
i  ([H_{0},B] f, f)\geq c_{1}
\| \langle x \rangle^{-1/2} \nabla^{\bot} f\|^2- c_2
\| \langle x \rangle^{-r} f\|^2, \quad r>1/2.
\]
This yields inequality  (\ref{eq:SM2}) with $K= \langle x \rangle^{-r} $ which implies  the radiation estimate.

Given $H$-smoothness of the operators $\langle x \rangle^{-r} $ for $r>1/2$ and
  $\langle x \rangle^{-1/2} \nabla^{\bot}$, we are in a position to prove
the asymptotic completeness (\ref{eq:10}). First, we replace wave operators (\ref{eq:8M})  by wave operators (\ref{eq:8J})  with non-trivial ``identifications"
$J_{\pm}$.  We choose $J_{\pm}$ as pseudo-differential operators 
\[
(J_{\pm} f)(x)= (2\pi)^{-d/2}\int_{{\mathbb R}^d} e^{i\langle x,\xi\rangle} {\bf j}_{\pm} (x,\xi) \hat{f}(\xi) d\xi 
\]
with  suitable symbols ${\bf j}_{\pm} (x,\xi) $ depending on a long-range potential $V(x)$ and on the sign of time $t$.  
Then the ``effective" perturbation $T_{\pm}:=HJ_{\pm}-J_{\pm}H_{0}$ is also a pseudo-differential operator with symbol
\[
 {\bf t}_{\pm} (x,\xi)= (-\Delta +V(x)- |\xi|^2) {\bf j}_{\pm} (x,\xi).
\]
Up to some  cut-off functions, the symbol ${\bf j}_{\pm} (x,\xi)$ is defined in terms of solutions $\varphi_{\pm}(x,\xi)$ (perhaps, approximate)  of the eikonal equation
\begin{equation}
2 \langle \xi, \nabla \varphi_{\pm}(x,\xi)\rangle + | \nabla \varphi_{\pm}(x,\xi)\rangle|^2 +V (x)=0, \quad \nabla=\nabla_{x}.
\label{eq:pdo2}\end{equation}
For example, in the case $\rho>1/2$, one can set  
\[
\varphi_{\pm}(x,\xi)=\pm 2^{-1} \int_{0}^\infty \big(V(x\pm\tau \xi)- V(\pm\tau \xi) \big)d\tau.
\] 

A notorious difficulty of this  construction is  that  the function $ \varphi_{\pm}(x,\xi)$ satisfies equation (\ref{eq:pdo2}) only away of a conical neighborhood of the direction $\hat{x}=\mp \hat{\xi}$ which necessitates a   cut-off  $ \sigma_{\pm} (\langle \hat{x},  \hat{\xi}\rangle)$ in the  angular variable. Here 
$ \sigma_{\pm} \in C_{0}^\infty (-1,1)$ is such that $ \sigma_{\pm} (\theta)=1$ in a neighborhood of the point $\pm 1$ and $ \sigma_{\pm} (\theta)=0$ in a neighborhood of the point $\mp 1$. Then we set
\[
 {\bf j}_{\pm} (x,\xi)= e^{i \varphi_{\pm}(x,\xi)} \sigma_{\pm} (\langle \hat{x},  \hat{\xi}\rangle).
\]

Combined together,  the limiting absorption principle and   the radiation estimate  show that the perturbation $T_{\pm}$ is a combination of $H_{0}$- and $H$-smooth operators. This implies the existence of the wave operators
\[
W^{(\pm)} (H,H_{0} ;J_{\pm})\quad \mbox{and} \quad W^{(\pm)} (H_{0} ,H;J_{\pm}^*).
\]
This result is an analytic background for the proof of 
the asymptotic completeness (\ref{eq:10}) as well as of the coincidence of the wave operators $W^{(\pm)} (H,H_{0} ;J_{\pm})$ and (\ref{eq:8M}). These wave operators are isometric. Thus, properties of the wave operators are quite similar in the short- and long-range cases. 

The scattering matrix $S(\lambda)$  is again defined by relations (\ref{eq:11}) and (\ref{eq:sm}),  but its properties are now quite different from the short-range case. Indeed, for short-range potentials, $S(\lambda)$  is close to the identity operator $I$;  more precisely, the operator $S(\lambda)-I$ is compact. Its integral kernel (the scattering amplitude) has only a weak diagonal singularity which however gets stronger if $V(x)$ decays slower.
On the contrary,  for long-range potentials,  the spectrum of $S(\lambda)$ covers the whole unit circle and the diagonal singularity of its integral kernel $s(\omega,\omega' ; \lambda)$  is quite wild:
\begin{equation}
s(\omega,\omega'; \lambda)\sim c(\omega;\lambda) |\omega-\omega'|^{-(d-1) (1+\gamma)/2}  e^ {ic_{0}(\omega;\lambda) |\omega-\omega'|^{ 1-\gamma}} 
\label{eq:diag}\end{equation}
where $\gamma=1/\rho$;
we refer to \S 9.5 in \cite{LNM} for more details.  Thus, the differential cross-section is proportional to
\[
|s(\omega,\omega'; \lambda)|^2\sim |c(\omega;\lambda)|^2  |\omega-\omega'|^{-(d-1) (1+\gamma)}  ,
\]
which yields a generalization of the Gordon and Mott (or Rutherford) formula  for the Coulomb potental.  Since $(1+\gamma)/2>1$, the diagonal singularity of $s(\omega,\omega'; \lambda)$ is stronger than the singular denominator. Nevertheless,  the operator $S(\lambda)$ with this kernel is bounded due to an oscillating factor in (\ref{eq:diag}).


We refer to  Chapter~11 of  the book  \cite{Y} for
a more detailed presentation of  long-range scattering theory.
\section{Scattering in three-particle systems}

\subsection{Setting the problem} 

Let us now consider three-particle systems described by Hamiltonians
\[
{\mathbf H}= {\mathbf H}_{0}+ V \quad \mbox{where}\quad V=V_{12}+ V_{23}+ V_{31}
\]
acting in the space
$L^{2} ({\mathbb R}^{3d})$. Here
\[
 {\mathbf H}_{0} = -(2m_{1})^{-1}\Delta_{x_{1}} -(2m_{2})^{-1}\Delta_{x_{2}} -(2m_{3})^{-1}\Delta_{x_{3}}, \quad x_{j}\in {\mathbb R}^{d},
\]
    is the kinetic energy operator   and $V_{\alpha}$ are the operators  of interactions of pairs  $\alpha=(12), (23), (31)$. They act as multiplications by functions (potential energies of pair interactions); for example,
    \begin{equation}
(V_{12} f) (x)= V_{12}(x_{1}-x_{2}) f(x),\quad x=(x_{1}, x_{2}, x_3). 
\label{eq:2.3}\end{equation}
  Expressions for other operators $V_{\alpha}$ are obtained from  (\ref{eq:2.3}) by cyclic permutations of indices. We usually suppose that $d=3$.
  
   The operator ${\mathbf H}$ commutes with shifts $x_{j}\mapsto  x_{j}+ x_{0}$ where $j=1,2,3$ and $x_{0}\in {\mathbb R}^d$ is an arbitrary vector (thus,  ${\mathbf H}$
is translationally invariant). This allows one to remove the center-of-mass motion and to reduce ${\mathbf H}$ to an operator acting in the space ${\cal H}=L^2 ({\mathbb R}^{2d})$.  The reduced Hamiltonian  is again given   by  the equality
$H= H_{0} + V$  where the potential energy $V$ is  defined   by formulas of type  (\ref{eq:2.3})  and the kinetic energy is the differential operator
\[
H_{0} = -(2m_{\alpha})^{- 1}\Delta_{x_{\alpha}} -(2n_{\alpha})^{- 1}\Delta_{y_{\alpha}}. 
\]
Here $\alpha$ is any of the pairs $(12), (23), (31)$, $m_{\alpha}$, $n_{\alpha}$ are reduced masses,
\[
m_{12}^{-1}=m_{1}^{-1}+ m_{2}^{-1}, \quad n_{12}^{-1}= (m_{1} + m_{2})^{-1} + m_{3}^{-1},
\]
and $ x_{\alpha}$, $y_{\alpha}$ are the Jacobi coordinates.  For example,
\[
x_{12}= x_{1}- x_{2}, \quad y_{12}= x_3 - (m_{1}+ m_{2})^{-1} (m_{1}x_{1}+ m_{2}x_{2}).
\]

We suppose that
pair potential energies
$V_{\alpha} (x_{\alpha})$ 
are bounded real functions on ${\mathbb R}^d$ and that they tend to zero sufficiently rapidly (more rapidly than the Coulomb potential) as $|x_{\alpha}|\to\infty$.  However for three-particle Hamiltonians, the total potential energy $V(x)$ does not tend to zero as $|x|\to\infty$ even after the separation of the center-of-mass motion. Indeed, suppose that two particles, for example, the first and the second, remain close to each other while the third one is far away. Then the functions $V_{23}(x_{2}-x_{3})$ and $ V_{31} (x_{3}-x_{1})$ tend to zero as $|x|\to\infty$ in ${\mathbb R}^{2d}$, but this is not true for $V_{12}(x_{1}-x_{2})$.  Thus, $V(x)$ does not tend to zero as   $|x|\to\infty$ in the configuration space ${\mathbb R}^{2d}$ of three-particle systems. This is the crucial difference with the two-particle case which leads to new scattering channels.

Nevertheless using, for example,  Cook's criterion, one can easily show that the wave operators $W_{0}^{(\pm )} : =W^{(\pm )} (H, H_{0})$ exist.  They describe the channel of scattering where
all particles are asymptotically free for $t\to\pm\infty$. This means that if a vector $f$ belongs to the range ${\rm Ran}\: W^{(\pm )}_0$ of the wave operator $W^{(\pm )}_0$, then
\[
\lim_{t\to\pm \infty}\| e^{-i H t} f- e^{-i H_{0}t} f_{0}^{(\pm)}\|=0
\]
where $f_{0}^{(\pm)}=\big( W^{(\pm )}_0 \big)^*f$.

Additionally, for three-particle systems   may exist  channels   corresponding to  scattering of one of the particles  on a bound state of two other particles. In this case the completeness of the wave operators $W_{0}^{(\pm )}  $ is lost.  To describe additional  channels, on has to introduce two-particle Hamiltonians
\begin{equation}
H^{\alpha}= - ( 2m_{\alpha})^{-1}\Delta + V_{\alpha}
\quad\mbox{and}\quad K_{\alpha}= -(2n_{\alpha})^{-1}\Delta  
\label{eq:2.5} \end{equation}
acting in the space $ L^2 ({\mathbb R}^d)$.
Suppose that the operator $H^{\alpha}$ has   negative eigenvalues $\lambda_{\alpha,n}$ where     $n=1,2,\dots, N_{\alpha}$ (the case $N_{\alpha}=\infty$  
is not excluded) with the corresponding normalized eigenfunctions $\psi_{\alpha,n}(x)$.  For every $\alpha$ and $n$, we put $a=(\alpha,n)$  and consider the evolution $U_a (t): L^2 ({\mathbb R}^{d})\to L^2 ({\mathbb R}^{2d})$ defined by the formula
\[
(U_a (t) f)(x)= \psi_a (x_{\alpha}) (e^{-i (K_\alpha+\lambda_{a}) t} f)(y_{\alpha})  .
\]
Using  again Cook's criterion, one can show that the limits
\begin{equation}
W^{(\pm )}_a  = \mbox{s-}\!\!\!\lim_{t\to\pm\infty}\exp(i  H t)
U_a (t)
\label{eq:2.7}\end{equation}
exist. The operators $W^{(\pm )}_a :L^2 ({\mathbb R}^d) \to :L^2 ({\mathbb R}^{2d})$ are isometric, enjoy the intertwining property $H  W^{(\pm )}_a=W^{(\pm )}_a (K_\alpha+\lambda_{a})$ and, for different $\alpha$ or $n$,  their ranges ${\rm Ran}\: W^{(\pm )}_a$ are orthogonal to each other   and to the subspace ${\rm Ran}\: W_{0}^{(\pm )}  $.  The  difficult result is {\it the asymptotic completeness}:
 \begin{equation}
 {\rm Ran}\:  W^{(\pm )}_{0}\oplus \Big(\bigoplus_{\alpha,n}  {\rm Ran}\: W^{(\pm )}_{\alpha,n} \Big)={\cal H}_{c}.  
\label{eq:2.8}\end{equation} 
 Thus  for every vector $f\in {\cal H}_{c}$, the evolution of the system governed by the Hamiltonian $ H $ decomposes 
asymptotically into a sum of  simpler evolutions determined by subsystems:
\begin{equation}
 \lim_{t\rightarrow\pm\infty}\| \exp(-i H t)f- \exp(-i H_{0}t)f_{0}- \sum_{\alpha,n}  U_{\alpha,n} (t) f_{\alpha,n}
 \|=0 
\label{eq:2.9}\end{equation}
where $f_{0}= (W^{(\pm)}_{0})^{*}f \in L^2 ({\mathbb R}^{2d})$ and $f_{\alpha,n}=(W^{(\pm )}_{\alpha,n})^{*} f \in L^2 ({\mathbb R}^d)$. The asymptotic completeness implies that no other scattering channels are  possible.
  This statement is  almost obvious physically, but it is hard to rigorously prove.  More than that, it is even false if pair potentials decay sufficiently slowly at infinity -- see Sect.~2.6.  From the  physics point of view,  relation (\ref{eq:2.9})  means that
 every bound state of each pair of particles asymptotically  behaves    as a new particle.
 
 Note that the spectra of all operators $H_{0}$ and $K_\alpha$ are (absolutely) continuous and coincide with the half-line $[0,\infty)$. It follows from (\ref{eq:2.8}) and the intertwining property of the wave operators that the continuous spectrum of the Hamiltonian $H$ is the union of the intervals $[0,\infty)$ and $[\lambda_{\alpha,n},\infty)$ over all $\alpha,n$.

 It is convenient to state the above results in slightly different terms introducing three-particle operators
 \begin{equation}
 H_{\alpha}= H_{0}+ V_{\alpha}= H^\alpha + K_{\alpha}
\label{eq:2.10}\end{equation}
 with only one pair interaction $V_{\alpha}$. The operator $H_{\alpha}$ is simply described in the Jacobi coordinates $(x_{\alpha}, y_{\alpha})$  of the space ${\mathbb R}^{2d}$:  
\begin{equation}
( H_{\alpha} f)(x_{\alpha}, y_{\alpha})= ( H^{\alpha} f)(x_{\alpha})+ ( K_{\alpha} f)( y_{\alpha}),
\label{eq:2.10a}\end{equation}
where the operators $H^{\alpha} $ and $K_{\alpha} $  are given by equations (\ref{eq:2.5}) in the variables $x_{\alpha}$ and $y_{\alpha}$, respectively.
 Consider   decomposition (\ref{eq:dc}) of the space ${\cal H}^\alpha $ into the orthogonal sum of the discrete ${\cal H}^\alpha_{p} $
  and continuous  ${\cal H}^\alpha_{c}$  subspaces of the operator $H^\alpha$, 
  and let $P^\alpha$ be the orthogonal projector in $L^2 ({\mathbb R}^d)$ onto the subspace ${\cal H}^\alpha_p$. Let us set $P_{\alpha}=P^\alpha\otimes I$ so that 
 \[
 (P_{\alpha} f) (x_{\alpha}, y_{\alpha})=\sum_{n=1}^{N_{\alpha}}\psi_{\alpha,n} (x_{\alpha})
 \int_{{\mathbb R}^d}\overline{\psi_{\alpha,n} (x_{\alpha}')} f(x_{\alpha}',y_{\alpha} ) dx_{\alpha}'.
 \]
 We   define  the   wave operators
 $
  W^{(\pm )} ( H , H_{\alpha};  P_{\alpha})
 $
 by formula  (\ref{eq:8J}) where $J_{\pm}= P_\alpha$. Note that the operators $H_{\alpha}$ and $P_{\alpha}$ commute. These wave operators exist and are isometric on the subspace $ {\rm Ran}\:  P_{\alpha}$. Their ranges are mutually orthogonal and the asymptotic completeness holds:
  \begin{equation}
    {\rm Ran}\:  W^{(\pm )} ( H , H_0  ) 
   \oplus \Big(\bigoplus_{\alpha}      {\rm Ran}\:  W^{(\pm )} ( H , H_{\alpha};  P_{\alpha}) \Big)={\cal H}_{c}.  
 \label{eq:asy} \end{equation}
 This is equivalent  to (\ref{eq:2.8}).

Below,  we briefly describe two different approaches to proofs of the asymptotic completeness. The first of them (Sect.~2.3 and 2.4) relies on a study of the resolvent of the three-particle Hamiltonian $H$. Its important advantage is that it  allows one to obtain convenient formulas for main objects of the  theory. The  second approach  (Sect.~2.5) is more abstract. It avoids a detailed study of the resolvent and can be directly generalized to an arbitrary number of particles.  Analytically, this second approach has some common features with the method used in Sect.~1.5 for long-range potentials.

\subsection{Separation of variables} 

Here, we consider a toy problem where the three-particle scattering theory reduces to the two-particle case. Suppose that the mass of one of the particles is infinite, for example $m_{3}=\infty$, and that the first and second  particles do not interact with each other, that is, $V_{12}=0$.
Let us fix a position of the third particle as $x_{3}=0$.  Then
 \[
H=H^{(1)}  + H^{(2)}
\]
  where  the two-particles operators
\[
H^{(1)} = -(2m_{1})^{-1}\Delta_{x_{1}}  + V_{31}(-x_{1}) \quad \mbox{and}\quad H^{(2)} =-(2m_{2})^{-1}\Delta_{x_{2}}   + V_{23} (x_{2} )
\]
act in the space $L^2 ({\mathbb R}^d)$ in the variables $x_{1}$ and $x_{2}$,  respectively. 
 For $j=1,2$, consider decomposition (\ref{eq:dc}) of $L^2 ({\mathbb R}^d_{x_{j}})$ into the orthogonal sum of the ``discrete" ${\mathcal H}^{(j)}_{p}$  and  ``continuous" ${\mathcal H}^{(j)}_{c}$ subspaces.
Then
\begin{eqnarray}
L^2 ({\mathbb R}^{2d})&=&L^2 ({\mathbb R}^d_{x_{1}})\otimes L^2 ({\mathbb R}_{x_{2}}^d)
\nonumber\\
=
\big({\mathcal H}^{(1)}_{c} \otimes {\mathcal H}^{(2)}_{c}\big)
&\oplus &\big( {\mathcal H}^{(1)}_{p} \otimes {\mathcal H}^{(2)}_{c}\big)
\oplus \big( {\mathcal H}^{(1)}_{c} \otimes {\mathcal H}^{(2)}_{p}\big)
 \oplus \big({\mathcal H}^{(1)}_{p} \otimes {\mathcal H}^{(2)}_{p}\big).
\label{eq:sep}\end{eqnarray}
The orthogonal projector in the space ${\mathcal H}^{(j)}$ onto ${\mathcal H}^{(j)}_{p}$ will be denoted $P^{(j)}$.

 Since the operators $H^{(1)} $ and $H^{(2)} $  commute, we have
\begin{equation}
e^{ i  H t}= e^{ iH^{(1)}t} \otimes e^{ i H^{(2)}t},
\label{eq:sep1}\end{equation}
whence
\begin{equation}
W^{(\pm)} ( H , H_0)= W^{(\pm)} (H^{(1)} , H^{(1)}_{0}) \otimes W^{(\pm)}  (H^{(2)} , H^{(2)}_{0})
\label{eq:sep2}\end{equation}
(the tensor product simply means that the operators act in different variables)
where $H^{(j)}_{0}= -(2m_{j})^{-1}\Delta_{x_{j}} $, $j=1,2$, are the kinetic energy operators.  The two-particle asymptotic completeness implies that $ {\rm Ran}\:  W^{(\pm)} (H^{(j)} , H^{(j)}_0) = {\mathcal H}^{(j)}_{ c} $ and therefore by virtue of (\ref{eq:sep2})
\[
 {\rm Ran}\:  W^{(\pm)} ( H  , H _0) = {\mathcal H}^{(1)}_c\otimes {\mathcal H}^{(2)}_c.
 \]
 
 Following (\ref{eq:2.10a}), we put $H_{j}= H^{(j)}  + H_{0}^{(k)}$ where $k\neq j$.
According to  (\ref{eq:sep1})   
\begin{eqnarray*}
e^{i    H t}   e^{- i    H_j t}& = &e^{ i(H^{(j)}+H^{(k)})t} e^{- i   ( H^{(j)}  +    H^{(k)}_{0}) t} 
\\
&=& I \otimes e^{ i H^{(k)} t} e^{- i   H^{(k)}_{0} t} \to  I \otimes   W^{(\pm )} (   H^{(k)}, H^{(k)}_{0})
 \end{eqnarray*}
as $t\to\pm \infty$ whence
\[
W^{(\pm )} ( H , H_j)= I \otimes W^{(\pm )} ( H^{(k)},  H^{(k)}_{0}).
\]
The completeness of the wave operator $ W^{(\pm)} (H^{(k)} , H^{(k)}_{0})$ ensures that 
\[
 {\rm Ran}\:   W^{(\pm )} ( H , H_j ;  P_j)=  {\cal H}^{(j)}_{p} \otimes  {\cal H}^{(k)}_{c}.
\]

 It now follows that the left-hand side of (\ref{eq:asy})  equals the orthogonal sum of the first three terms on the right in (\ref{eq:sep}). Since
 ${\mathcal H}_{p}={\mathcal H}^{(1)}_{p} \otimes {\mathcal H}^{(2)}_{p}$, the asymptotic completeness (\ref{eq:asy}) follows.
 

\subsection{Resolvent equations} 

In the three-particle case, the resolvent equation (\ref{eq:14}) is not Fredholm  even   for $\Im z\neq 0$. To overcome this difficulty, Faddeev split (see his monograph \cite{Fadd}) the resolvent $R(z) =(H-zI)^{-1}$ of the operator $H=H_{0}+ \sum V_{\alpha}$ into three components $T_{\alpha}(z)=  \sqrt{|V_{\alpha}|} R (z)$   and derived a Fredholm system of equations for them.
 The entries of this system are constructed in terms of three Hamiltonians $H_{\alpha}$ defined by equality (\ref{eq:2.10}).
    To derive a system for the operators  $T_{\alpha}(z)$, we
  write down the resolvent equation
\[
R(z)= R_{\alpha}  (z) -R_{\alpha}  (z)  \sum_{\beta\neq\alpha}  V_{\beta} R(z), \quad R_{\alpha}  (z)= (H_{\alpha}-zI)^{-1},
\]
for each pair $H_{\alpha}, H$. Then we multiply it  by $\sqrt{|V_{\alpha}|}$ and set 
\begin{equation}
T_{\alpha}^{(0)}(z)=  \sqrt{|V_{\alpha}|} R_{\alpha} (z), \quad Q_{\alpha,\beta} (z)=    \sqrt{|V_{\alpha}|} R_{\alpha} (z) \sqrt{V_{\beta}} 
\label{eq:2.11A}\end{equation}
 (here $\sqrt{V_{\alpha}} =V_{\alpha} \sqrt{|V_{\alpha}|}^{-1}$).
 This yields a system of three equations (Faddeev's equations)
\begin{equation}
T_{\alpha}(z) = T_{\alpha}^{(0)}(z)  -  \sum_{\beta\neq\alpha} Q_{\alpha,\beta} (z) T_{\beta} (z), \quad \Im z\neq 0,
\label{eq:2.12}\end{equation}
for three operators $T_{\alpha}(z)$. 

{\it The crucial point is that the operators $Q_{\alpha,\beta} (z)$ are compact because, for $\beta\neq\alpha$,  the products $V_{\alpha}(x_{\alpha} )V_{\beta} (x_{\beta})$  tend to zero as $|x|\to\infty$ in } ${\mathbb R}^{2d}$.  Moreover, the homogeneous system
\begin{equation}
f_{\alpha} =   -  \sum_{\beta\neq\alpha} Q_{\alpha,\beta} (z) f_{\beta}  ,\quad \alpha=(12), (23), (31),
\label{eq:comp}\end{equation}
 corresponding to (\ref{eq:2.12}) has only a trivial solution $f_{12}=f_{23}= f_{31}=0$. This is a consequence of the self-adjointness of the operator $H$. Thus, by the Fredholm alternative,  the non-homogeneous system (\ref{eq:2.12}) has a unique solution $T_{12}(z),T_{23}(z)$, $T_{31}(z)$.  For every $\alpha$, the resolvent $R(z)$ can be recovered by the formula
\[
R(z)= R_{\alpha}  (z) -R_{\alpha}  (z)  \sum_{\beta\neq\alpha} \sqrt{V_{\beta}}   T_{\beta} (z).
\]
We emphasize that this equation contains  the resolvents $R_{\alpha}  (z) $ of all subsystems.

A passage to the limit  $\Im z\to 0$  requires additional considerations.    We suppose now that all pair potentials $V_{\alpha}$ satisfy condition  (\ref{eq:4})    for some $\rho>2$ and $d\geq 3$.  Then the two-particle operators $H^\alpha$ defined by formula  (\ref{eq:2.5}) in the space  $L^2({\mathbb R}^d)$ may have only finite numbers of eigenvalues. First, we discuss the case where they do not have eigenvalues at all.  Under some technical assumptions this implies that  their resolvents $R^\alpha (z)=(H^\alpha-z I)^{-1}$   sandwiched  by the operators 
$\sqrt{|V_{\alpha}|} $ are continuous (always  norm-continuous)  in $z$ up to the cut along $[0,\infty)$.  Using relation (\ref{eq:2.10a}), one can deduce from this result that the operator-valued functions $\sqrt{|V_{\alpha}|} R_{\alpha} (z) \sqrt{|V_{\alpha}|} $ acting in the space $L^2({\mathbb R}^{2d})$ possess the same property.  This is an essentially two-particle result.

A genuinely three-particle analytic result concernes the free resolvent $R_{0}(z)$ sandwiched by the operators $\sqrt{|V_{\alpha}|} $  and $\sqrt{|V_{\beta}|} $  for $\alpha\neq \beta$.  It turns out that the operator-valued  functions $\sqrt{|V_{\alpha}|}R_{0}(z)\sqrt{|V_{\beta}|}$ are continuous in  $z$ as   it approaches the positive half-axis (the spectrum of the operator $H_{0}$).  Using   the facts stated above, one can check that the operator-valued  functions $G_{\alpha,\beta} (z) $ defined in (\ref{eq:2.11A})  possess  the same property.

Now we turn to system (\ref{eq:2.12}).  Similarly to the two-particle case (cf. equation (\ref{eq:res2})),
 the results on entries of this system imply
  that, for all $\alpha$ and $\beta$, 
the operator-valued  functions $ T_{\alpha} (z)\sqrt{|V_{\beta}|}= \sqrt{|V_{\alpha}|}R (z)\sqrt{|V_{\beta}|}$ are also  continuous in  $z$ as it tends to the half-line $[0,\infty)$.
 
This result   shows that all operators $\sqrt{|V_{\alpha}|}$ are $H_{0}$- and $H$-smooth on ${\mathbb R}_{+}$. As explained in Sect.~1.3, this ensures  that the wave operators $W_{\pm} (H,H_{0})$ exist and are complete. It follows that the  scattering operator ${\bf S}$ defined by  relation  (\ref{eq:11})   and  the corresponding scattering matrix $S(\lambda)$, $\lambda\in{\mathbb R}_{+}$, are unitary operators in the spaces  $L^2 ({\mathbb R}^{2d})$ and $L^2 ({\mathbb S}^{2d-1})$, respectively.  Moreover, $S(\lambda)$ satisfies representation (\ref{eq:33}), and the Born expansion (\ref{eq:33A}) is also true.
The right-hand sides here are  well defined in view of analytic results listed above.  An important difference with the two-particle case is that now the operator $S(\lambda)-I$ is not compact. Nevertheless, it is possible to describe its structure in terms of the scattering matrices $S_{\alpha} (\lambda)$   for the pairs $H_{0}$, $H_{\alpha}$ where $\alpha=(1 2), (2 3)$, $(3 1)$. Indeed, it follows from (\ref{eq:33A}) that
\[
 S(\lambda)=I + \sum_{\alpha} (S_{\alpha}(\lambda)-I) + \widetilde{S}(\lambda)
\]
where the operator
\[
\widetilde{S}(\lambda)=
  -2\pi i \sum_{\alpha\neq \beta}\sum_{n=0}^\infty (-1)^n \Gamma_0(\lambda)V_{\alpha}(R_0(\lambda+i0)V_{\beta})^n \Gamma_0^\ast (\lambda).
\]
The operators $S_{\alpha} (\lambda)$ can be expressed in terms of scattering matrices for two-particle operators $H_{0}^{\alpha}$, $H^{\alpha}$.
The operator $\widetilde{S}(\lambda)$ is compact because the products $V_{\alpha} R_0(\lambda+i0)V_{\beta}$  are compact.  The operator $\widetilde{S}(\lambda)$ is known as the connected part of the scattering matrix $S (\lambda)$.  Note that its integral kernel does not contain the Dirac delta-functions.


\subsection{Multi-channel case}    


Let us finally comment the general case where the two-particles operators $H^\alpha$ defined by equality  (\ref{eq:2.5}) have   finite numbers of negative eigenvalues. To simplify notation, we assume that every $H^\alpha$ has exactly one negative eigenvalue $\lambda_{\alpha}$.

  Now we have to single out the poles of their resolvents $R^\alpha (z)=(H^\alpha-zI)^{-1}$   corresponding to these eigenvalues. Thus, we set
\begin{equation}
 R^\alpha (z)=\widehat{R}^\alpha (z)+  P^{\alpha} (\lambda_{\alpha}-z)^{-1}
\label{eq:33C1}\end{equation}
 where $P^{\alpha}= \psi_{\alpha} (\cdot, \psi_{\alpha})$ is the orthogonal projection in $L^2 ({\mathbb R}^d)$ on the normalized eigenfunction $\psi_{\alpha}$ of the operator $H^\alpha$.  In view of the pole  in (\ref{eq:33C1}) there are no chances that the sandwiched  resolvents $\sqrt{|V_\alpha |}  R^\alpha(z) \sqrt{| V_\beta |} $ and a fortiori $ \sqrt{|V_{\alpha}|}  R (z)   \sqrt{|V_\beta|} $ are continuous up to the continuous spectrum of the operator $H$. However such  a statement is true if the pole terms in (\ref{eq:33C1}) are removed. To be precise, using system (\ref{eq:2.12}), one can prove that the operator-valued functions
 \[
 \sqrt{| V_{\alpha}|}  (I - P_{\alpha}) R(z) (I - P_{\beta})  \sqrt{| V_\beta|}      ,
 \quad P_{\alpha}= P^\alpha \otimes I,
 \]
 are norm-continuous for all $\alpha$ and $\beta$   as $z$ approaches the cut along $[ E, \infty)$ where  $
E=\min_{\alpha}\{\lambda_{\alpha}\} $.  This implies that the operators $  \sqrt{|V_{\alpha}|}   (I - P_{\alpha}) $ are $H_{0}$- and $H$-smooth, and hence the wave operators 
\begin{equation}
W^{(\pm)} (H,H_{0}; I-\sum_{\alpha}P_{\alpha})\quad \mbox{as well as} \quad W^{(\pm)} (H_{0},H; I-\sum_{\alpha}P_{\alpha})
\label{eq:33C2}\end{equation}
exist. 
With significant simplifications, the above arguments prove also the existence of the wave operators 
\begin{equation}
W^{(\pm)} (H,H_{\alpha};  P_{\alpha} )\quad \mbox{and} \quad W^{(\pm)} (H_{\alpha}, H;  P_{\alpha} )
\label{eq:33C23}\end{equation}
  corresponding to two-particle scattering channels.  
  
  Note that, for all $\alpha$,  
\[
  \mbox{s-}\!\!\!\lim_{|t|\to\infty} P_{\alpha}\exp(-iH_{0}t)
=0
\]
and that the  operators $P_{\alpha} P_{\beta}$ are compact if $\alpha\neq \beta$.  Therefore the asymptotic completeness (\ref{eq:asy}) can be deduced from the results on wave operators (\ref{eq:33C2}) and (\ref{eq:33C23}) stated above.


\subsection{Non-perturbative approach}    


 The approach described briefly in two previous subsections relies on a kind of an advanced perturbation theory where  the free problem is determined by the set of all sub-Hamiltonians. Its generalization to the case of an arbitrary number of particles meets with serious difficulties. Here we discuss a different, non-perturbative, approach which works well for any number of particles  of arbitrary dimension $d$.
In the non-perturbative approach   the operators $ H$ and $H_{0}$ as well as the Hamiltonians $H_{\alpha}=H_{0} +V_{\alpha}$  of all subsustems are treated on an  equal basis. It is supposed that all pair potentials satisfy condition (\ref{eq:4}) for some $\rho>1$. No assumptions on subsystems are required.  Below the index $a=0$ or $a=\alpha$ where as usual $\alpha=(12), (23), (31)$.
 
 The idea of the proof of the asymptotic completeness is to find identifications $J_{\alpha}$  that ``kill" directions in ${\mathbb R}^{2d}$ where the potentials $V_{\alpha} (x_{\alpha})$ do not tend to zero as $|x|\to\infty$.  Then, for all triples $H_{a}$, $H$, $J_{a}$,  ``effective"
perturbations $H J_{a} - J_{a}  H_{a}$ consist of combinations of $H$- and $H_{a}$-smooth terms.
  It is convenient to choose $J_{a}$ as first-order differential  rather than multiplication operators. Unfortunately, the commutators $[H_{0}, J_{a} ]$ of $J_{a}$  with the operator $H_{0}$ have coefficients decaying at infinity   as $|x|^{-1}$ only.
 To a some extent, we follow the approach exposed in Sect.~1.5 which relies on  the limiting absorption principle and the radiation estimate (stating that the operator $ \langle  x \rangle^{-1/2} \nabla^\bot $ is $H$-smooth), but these analytic tools become essentially more complicated now. In particular, the Mourre estimate is satisfied only away from all eigenvalues  $\lambda_{\alpha, n}$ of subsystems (the thresholds of the operator $H$) and the operator-valued function $\langle  x \rangle^{-r} R(z) \langle  x \rangle^{-r} $ where $r>1/2$ is  continuous in $z$ only away from these points. 
 
 The two-particle radiation estimate  holds only in the part of the configuration space where all three particles are far away from each other.  In the region
 where $|x_{\alpha}|/ |x|$ is small,  the motion of the system is very complicated in the variable $x_{\alpha}$ but it is close to the free motion in the variable $y_{\alpha}$. In this variable, the radiation estimate remains true. To be precise, we denote by $\chi_{\alpha}  = \chi_{\alpha} (\varepsilon)$ the characteristic function of the cone  $\Gamma_{\alpha}=\Gamma_{\alpha}(\varepsilon)=\{ x: |x_{\alpha}| \leq \varepsilon |x| \}\subset{\mathbb R}^{2d}$  where $\varepsilon$ is sufficiently small. Then the operator 
 \[
G_\alpha= \chi_{\alpha}   \langle  x \rangle^{-1/2} \nabla^\bot_{y_{\alpha}}
 \]
 is $H$-smooth. This result is weaker than the corresponding statement in the two-particle case because $| (\nabla^\bot_{y_{\alpha}} f)(x) |  \leq | (\nabla^\bot f)(x) | $.  As was already mentioned, the operator 
 \[
G_0= \chi_{0}  \langle  x \rangle^{-1/2} \nabla^\bot \quad \mbox{ where } \chi_{0}  =1-\sum_{\alpha} \chi_{\alpha}
 \]
 is also $H$-smooth.

Similarly to the long-range case, our  proof of the radiation estimates is based on a consideration of the
commutator of $H$  with 
differential operator (\ref{eq:comm1}). Now $b(x)\neq |x|$, but it is close to $|x|$. An important difference with Sect.~1.5 is that now 
$b(x)$  depends only on $y_{\alpha}$ in cones $\Gamma_{\alpha}$  where we set
$b(x)= |y_{\alpha}|$.  The last circumstance ensures that the operators   $V_{\alpha}$ and $ B$ almost commute, that is, the commutators $[V_{\alpha},B]$ are short-range functions.   This allow us to prove, for all $a=\alpha$ and $a=0$,   estimates
\[
[H, B]\geq c\,  G_a^\ast G_a - c_{1}\langle
x\rangle^{-1-\varepsilon}, \quad \varepsilon>0, 
\]
which  imply that the operators  $G_a$ are $H$-smooth.

We introduce auxiliary wave operators
\begin{equation}
W^{(\pm)} (H, H_a; B_a  )\quad \mbox{and} \quad    W^{(\pm)} (H_a, H;   B_a  )
\label{eq:wo}\end{equation}
where  $B_a $ are  differential operators (\ref{eq:comm1}). Their coefficients $b_{a} (x)$ are smoothed   products $\chi_{a}(x) b (x)$.   Observe that  ``perturbations"
\begin{equation}
H B_{a}- B_{a}H_{a}=( V -V_{a}) B_{a}  + [ V_{a} , B_{a}] + [ H_{0} , B_{a}]
\label{eq:pert}\end{equation}
(we set $V_{0}=0$)
consist of combinations of smooth terms. Indeed, to consider  the first  term $( V -V_{a}) B_{a} $, it suffices to observe that all products $V_{\alpha}(x) \chi_{0}(x)$  as well as $V_{\beta}(x) \chi_{\alpha}(x)$ where $\beta\neq \alpha$  are short-range functions.  This fact is also true for  the second  term
$ [ V_{a} , B_{a}] $  because $m(x)= |y_{\alpha}|$ for $x\in \Gamma_{\alpha} $.
 Thus, the first two terms in the  right-hand side of (\ref{eq:pert})   can be taken into
account by the limiting absorption principle.   The commutator
$[H_0, B_a]$ factorizes into a product of
$H_a$- and $H$-smooth operators according to the radiation estimates.  This proves the existence of wave operators (\ref{eq:wo}). 

It remains to show that this fact implies the asymptotic completeness. Let us set
 \begin{equation}
 \sum_a b_a=b, \quad  B=\sum_a B_a
\label{eq:MM}\end{equation}
 and introduce   the wave operator
(observable)  
$
W^{(\pm)} (H, H ; \pm B  )  =: B_{\pm}    .            
$
 It exists by the
  arguments given above for  wave operators (\ref{eq:wo}).  As usual, we neglect cut-offs on energy intervals disjoint from eigenvalues and thresholds of the operator $H$. It is important that the operator $\pm B_{\pm} $ is positive definite,
  \begin{equation}
  \pm B_{\pm} \geq c>0,
\label{eq:M}\end{equation}
  whence its range
\begin{equation}
 {\rm Ran}\,  B_{\pm} ={\cal H} .
\label{eq:M1}\end{equation}
The proof of (\ref{eq:M}) relies on the inequality 
\begin{equation}
\| |x|  \exp(-iHt) f\|\geq c |t|  \|  f\|\
\label{eq:M2}\end{equation}
which can be deduced from
the Mourre estimate  (\ref{eq:mourre}). Inequality  (\ref{eq:M2}) means that the evolution $( \exp(-iHt) f)(x)$ essentially ``lives" outside of the ball $| x| \leq c |t|$.  This fact is physically quite natural because a quantum particle   asymptotically moves with a constant velocity.

  We can now conclude the (heuristic) proof of the asymptotic completeness. It follows from  (\ref{eq:M1}) that for all vectors $f\in  {\cal H}$ 
\begin{equation}
\lim_{t\rightarrow\pm\infty} \| \exp(-iHt)f -  B \exp(-iHt)f_{\pm}\|=0
\label{eq:M3}\end{equation}
 if  $f=B_{\pm }f_{\pm} $.  The existence of wave operators (\ref{eq:wo}) implies that for any $f_{\pm}$ and $f_{\pm}^{(a)} = W^{(\pm)} (H_a, H;   B_a  ) f_{\pm}$
\begin{equation}
\lim_{t\rightarrow\pm\infty}\| B_{a}\exp(-iHt)f_{\pm}  -\  \exp(-iH_a
t) f_{\pm}^{(a)}  \|=0.
\label{eq:M4}\end{equation}
Combining  relations (\ref{eq:M3}), (\ref{eq:M4}) and using the second equality (\ref{eq:MM}), we see that
 $\exp(-iHt)f$ satisfies relation (\ref{eq:2.9}), that is, it decomposes asymptotically into simpler evolutions 
$\exp(-iH_a t)f_{\pm}^{(a)} $.
This is one of the equivalent formulations of   the asymptotic completeness and leads to relation (\ref{eq:asy}).

Of course the above arguments are essentially intuitive. The precise proof can be found in \cite{Ast}.



 
   \subsection {Long-range interactions. New channels}

   The three-particle problem acquires a   long-range character if
pair potentials decay as   Coulomb potentials or slower. Similarly  to the two-particle problem (cf. (\ref{eq:8M1}), (\ref{eq:8M2}) and (\ref{eq:8M})),  for long-range potentials the definition  of wave operators $W^{(\pm)} (H,H_{0})$  and $W^{(\pm)}_{a}$ defined by (\ref{eq:2.7}) should be naturally modified. As in the short-range case, only the asymptotic completeness is a really difficult mathematical problem; see the book \cite{DG}.  Its proof requires an assumption
\[
  |(\partial^\kappa V_{\alpha} (x_{\alpha})| \leq C (1+|x_{\alpha} |)^{-\rho-|\kappa|} 
\]
for all $\alpha$, all multi-indices $\kappa$ and some $\rho >\sqrt{3}-1$. The last condition is very important. It allows one to prove relation (\ref{eq:2.9}) with asymptotic evolutions only slightly modified by phase factors.


On the contrary,  if pair potentials $V_{\alpha} (x)$ decay slower than
$|x_{\alpha}|^{-1/2}$, then the traditional picture of  scattering breaks down (see \S 15 of (\cite{LNM})). Actually, 
 a three-particle system may  have  additional scattering channels
   intermediary between the channel where all three particles are asymptotically free and the channels where a couple of particles form a bound state. In these additional  channels, the bound state of a couple of particles
depends on a position of the third particle, and it is destroyed asymptotically.


 We first consider the Schr\"odinger operator
 \[
H = -\Delta_{x_1} - \Delta_{x_2}  
+ V ( x_{1}, x_{2})  
 \]
 in the space $L^2 ({\mathbb R}^{2d})$
 with a sufficiently arbitrary potential  $V ( x_{1}, x_{2}) $ depending on both variables $x_{1}$ and $x_{2}$. Let us introduce a family of operators
 \[
 H(x_1)=- \Delta_{x_2} +V(x_1,x_{2})
 \]
 acting in the space  $L^2 ({\mathbb R}^{d})$  in the variable $x_{2 }$ and depending on the parameter $x_1 \in{\mathbb R}^{d}$.
 Suppose  that  $H(x_1)$
has an eigenvalue
$\lambda (x_1)$, and denote  the corresponding normalized eigenfunction by  $\psi (x_1,x_{2})$. Actually,
it suffices to assume that $\psi (x_1,x_{2})$ is an approximate eigenfunction, that is
\[
-  \Delta_{x_{2}}\psi (x_1,x_{2})+ V(x_1,x_{2})\psi (x_1,x_{2})=\lambda (x_1)\psi (x_1,x_{2})+Y(x_1,x_{2}),
 \]
where 
\[
\| Y(x_1 )\|_{L^2 ({\mathbb R}^{d}_{x_{2}})}=O (|x_1|^{-1-\varepsilon}), \quad \varepsilon>0, \quad{\rm as}\quad |x_1|\rightarrow\infty.
\]
In interesting situations the eigenvalue $\lambda (x_1)$ tends to zero as $|x_{1}|\to \infty$  
slower than $\vert x_1\vert ^{-1}$, that is,
\[
\lambda (x_1)\sim |x_{1}|^{-2\sigma}, \quad 2\sigma < 1. 
\]
Let us consider it as an ``effective" potential energy and associate (see Section~1.5) to  the
long-range potential 
$\lambda (x_1)$ the phase function  $ \Xi (x_1,t)$. Now  $ \Xi$
satisfies    eikonal equation (\ref{eq:Eik})
with $V(x)$ replaced by $\lambda(x_1)$, that is 
\begin{equation}
\partial \Xi /\partial t+|\nabla_{x_1} \Xi |^2+ \lambda(x_1)=0,
\label{eq:NC4ab}\end{equation}
perhaps, up to a term $O(|x_1|^{-1-\varepsilon})$.   Typically,  the
asymptotic behaviour of $\psi (x_1,x_{2})$ as $\lambda (x_1)\rightarrow 0$ has a certain
self-similarity:
\begin{equation}
\psi (x_1,x_{2}) \sim | \lambda (x_1)| ^{  d/4}\Psi(\sqrt{| \lambda (x_1)|  } x_{2}) 
 \label{eq:1.13}\end{equation} 
for some $\Psi\in L^{2}( {\mathbb R}^d)$.

  Let us introduce the operator of 	a modified free evolution $U_1(t)$ in the variable $x_{1}$ (cf. (\ref{eq:8M1})):
\begin{equation}
 (U_1(t)f_1 )(x_{1},x_{2})=  \psi (x_1,x_{2})\exp (i\Xi (x_1,t)) (2it)^{-d/2} \hat{f}_1(x_1/(2t)).
 \label{eq:NC1}\end{equation}
 Our goal is to show the existence of the limits
\begin{equation}
 \lim_{t\rightarrow\pm\infty} \exp (iHt)  U_1(t)f_{1}=: w^{(\pm)}f_{1}.
 \label{eq:NC2}\end{equation}
 These
  limits   exist (cf. Cook's criterion in Sect.~1.3) provided
\begin{equation}
\int_{-\infty}^\infty \| (i\partial/\partial t -H)U_1(t) f_1 \|dt <\infty \quad \mbox{for }\quad \hat{f}_1\in
C_0^\infty ({\Bbb R}^{d}\setminus \{0\}),
 \label{eq:NC3}\end{equation}
that is,  $U_1(t) f_1$ is a reasonably good approximate solution of the time-dependent Schr\"odinger
equation.  Unfortunately, due to self-similarity (\ref{eq:1.13})  the integrand in (\ref{eq:NC3}) behaves as $|t|^{-1}$ for
$|t|\rightarrow\infty$, so condition (\ref{eq:NC3}) is never satisfied. We verify  however  
this condition  for a   evolution $\widetilde{U_1}(t)$, 
\[
(\widetilde{U_1}(t)f_1)(x_1,x_{2})= \exp(i\sigma (4t)^{-1}|x_{2}|^2) ( U_1 (t)f_1)(x_1,x_{2}), 
\]
containing an additionl phase factor compared to (\ref{eq:NC1}).
This implies the existence of limits (\ref{eq:NC2}) with $U_1$ replaced by $\widetilde{U}_1$. Since, moreover, 
\[
\lim_{t\rightarrow\pm\infty} \|\widetilde{U}_1(t) f_1-U_1(t) f_1 \|=0, \quad \hat{f}_1\in C_0^\infty
({\Bbb R}^{d}\setminus \{0\}),
\]
the wave operators $w^{(\pm)}$ also exist. Thus, for every
$f\in {\rm Ran}\,  w^{(\pm)} $ and $f= w^{(\pm)} f_{1}$, we have
\begin{equation}
(\exp(-iHt) f )(x)= \psi (x_1,x_{2})\exp (i\Xi (x_1,t)) (2it)^{-d/2} \hat{f}_1(x_1/(2t))+ \varepsilon (x,t)
 \label{eq:NCx}\end{equation}
where $\varepsilon (\cdot,t)\to 0$ in $L^2 ({\mathbb R}^{2d})$. The wave  operators $w^{(\pm)}: L_2({\mathbb R}^{d})\to L_2({\mathbb R}^{2d})$ are    isometric and  the intertwining property $H w^{(\pm)}=w^{(\pm)} (-\Delta_{x_1})$ holds. Their ranges are orthogonal to the ranges of the wave operators $W_{0}^{(\pm)} $  and $W_{\alpha}^{(\pm)} $.
According to (\ref{eq:1.13})  functions (\ref{eq:NC1}) ``live" for large $|t|$ in the region where $|x_{1}|\sim |t|$ and $|x_{2}| \sim |x_1|^\sigma\sim
|t|^\sigma$.  Relation (\ref{eq:NCx})  implies that the same is true for $\exp(-iHt)f$ if $f\in  {\rm Ran}\, w^{(\pm)}$.   


If $V(x_1,x_{2})=V( x_{2})$, then $H$ is the three-particle Hamiltonian with only one non-trivial pair
interaction. In this case $\psi(x_1,x_{2})=\psi( x_{2})$ is an eigenfunction of this pair, $\lambda$
does not depend on $x_1$ and
$w^{(\pm)}$ describes scattering of the third particle on this bound state. In the general case, $H$ can
be considered as the three-particle Hamiltonian with potential energy depending sufficiently
arbitrarily on positions of particles. Then  $w^{(\pm)}$ describes a channel where a pair of particles
is in a bound state  depending on the position of the third particle, and  the third particle is
being scattered by the ``effective" potential  $\lambda(x_1)$.

 Let us now come back to   three-particle Hamiltonians  with non-trivial pair interactions.
 For simplicity of notation, we assume that $m_{1}=m_{2}= 1/2$, $m_{3}=\infty$ and  $V_{31}=0$.
 We fix the position of the third particle by the condition $x_{3}=0$. Then
\[
H = - \Delta_{x_1} - \Delta_{x_2}  
+ V_{12}( x_{1}-x_{2}) + V_{23}( x_{2} ) .
 \]
 We assume that the pair potential  $V_{12}$  is long-range, 
  \begin{equation}
 V_{12}( x_{1}-x_{2} ) = v_{12}| x_{1}-x_{2} |^{-\rho}, \quad  v_{12} >0, \quad \rho<1/2,
 \label{eq:nc2a}\end{equation}
 for sufficiently large positive $ x_{1}-x_{2}$
 and $ V_{23}( x_{2} ) $  is short-range  and repulsive.  We also suppose that particles are one-dimensional, that is, $d=1$.  Then we use that for large $x_{1} > x_{2}\geq 0$, potential (\ref{eq:nc2a}) is well approximated by the linear function 
   \begin{equation}
  v_{12} \big( | x_{1}|^{-\rho} +    | x_{1}|^{-\rho-1} x_{2}\big)
 \label{eq:nc2b}\end{equation}
 of the variable $x_{2}$. This allows us to construct the function $\psi(x_1, x_{2})$ by the following procedure. Let $\Lambda$ be any eigenvalue and $\Psi  $ be the corresponding eigenfunction (the Airy function) of the equation
 \[
 -  \Psi'' (x_{2}) + v_{12} \rho x_{2} \Psi (x_{2}) = \Lambda\Psi (x_{2}), \quad x_{2 }\geq 0, \quad \Psi (0)=0,
 \]
 extended by zero to $x_{2}\leq 0$. Define the function  $\psi(x_1, x_{2})$ and the ``potential $ \lambda(x_{1})$ by the equations 
 \[
 \psi(x_{1}, x_{2})= |x_{1}|^{-\sigma/2}\Psi ( |x_{1}|^{-\sigma}x_{2}), \quad \sigma=(\rho+1)/3,
 \]
 \[
 \lambda(x_{1})=v_{12} |x_{1}|^{-\rho}+ \Lambda |x_{1}|^{- 2\sigma},
 \]
 and let $\Xi (x_1,t)$ be the phase function associated to the long-range potential $ \lambda(x_{1})$, that is,  $\Xi (x_1,t)$ is an (approximate) solution of equation (\ref{eq:NC4ab}). Let the  evolution $U_{1}(t)$ be defined by formula  (\ref{eq:NC1}).
 Then the wave operator $w^{(\pm)}: L^2 ({\mathbb R}_{\pm}) \to L^2 ({\mathbb R}  )$ defined by equality (\ref{eq:NC2})  exists.
 
Note that for $f\in  {\rm Ran}\,  w^{(\pm)} $, the solution $u(t)=\exp (-iH t) f$ of the Schr\"odin\-ger equation (\ref{eq:1})  ``lives" for   $t\to \pm \infty$ in the region  where  $x_{1}\sim |t|$ and $x_{2}\sim |t|^\sigma$ for $\sigma\in (1/3, 1/2)$. Such solutions describe a physical process where the second particle is relatively close to the third one and the first particle is far away. The pair $ (23)$ is  bound by a potential depending on the position of the first particle, but this bound state is evanescent as $|t|\to\infty$.
One  can imagine, for example, that the first particle moves to $+\infty$ and $x_{1}$ grows as $|t|$ as  $|t|\to\infty$, the second particle  is jammed to the origin by a repulsive interaction with the first particle. In the long run, it escapes to $+\infty$ but $x_{2} \sim |t|^\sigma$ for large $|t|$.

Similar phenomenon  occurs if $v_{12}<0$ in  (\ref{eq:nc2a}). In this case the first particle moves ``freely" to $-\infty$, and the second one  is located in the potential well (\ref{eq:nc2b}).

 
   \subsection {Discrete spectrum}

Here we suppose that that the dimension of particles  $d=3$. Recall that, in the two-particle systems,  the negative spectrum emerges if 
 the interaction $V(x)$  has a sufficiently large  negative part. It is infinite if $V(x)$ tends to zero  slower than $|x|^{-2}$. For example, this is the case for the Coulomb negative  potential. Otherwise, the discrete spectrum of the operator $- (2m)^{-1}\Delta+ V(x)$ is finite. 
 
 Let us now  discuss three-particle Hamiltonians $H$.  The crucial difference between the cases of two and three particles is that, for three-particle  systems, the potential energy does not tend to zero even after the separation of the center-of-mass motion. Let  the operator $H^\alpha$  be defined by equality (\ref{eq:2.5}), let  $\lambda_{\alpha,1}$ be its lowest eigenvalue (we set $\lambda_{\alpha,1}=0$  if $H^\alpha\geq 0$)  and
   \begin{equation}
E=\min_{\alpha}\{\lambda_{\alpha,1}\}.
  \label{eq:HVZ}  \end{equation}
If $H^\alpha\geq 0$, for all $\alpha$, we set $E=0$. 
It follows from the asymptotic completeness (\ref{eq:2.8}) that the essential spectrum of the three-particle Hamiltonian $ H$ coincides with the half-axis $[E,\infty)$.  This assertion (known as the Hunziker-Van Winter-Zhislin theorem) is true for all pair potentials $V_{\alpha} (x_{\alpha})$ decaying at infinity and naturally  extends  to an arbitrary number of particles.

We here briefly discuss the discrete spectrum of three-particle systems lying below the bottom $E$ of the essential spectrum. 
  In the case $E <0$, the situation is, roughly speaking, similar to two-particle  systems  (see  \cite{Y-Izv}  and \S 3.9 in the book  \cite{CFKS}). For pair potentials $V_{\alpha} (x_{\alpha})$ decaying faster than $|x_{\alpha}|^{-2}$,   three-particle Hamiltonians $  H$  may have only finite discrete spectra.  It is also possible that long-range pair potentials compensate each other. Consider, for example, the ion of the hydrogen atom  consisting of one proton and two electrons.  Now the bottom of the essential spectrum is determined by the lowest bound state of  the hydrogen atom. The extra electron interacts with the atom by the Coulomb potentials. One of them is attractive and another repulsive so  that the effective interaction of the ``free" electron with the atom is short-range. Therefore the hydrogen ion has only a finite number of bound states.  One of possible proofs of these results given in  \cite{Y-Izv} relies on the compactness of the operator $Q_{\alpha, \beta} (E)$ defined by formula (\ref{eq:2.11A}).
  
  The situation is drastically different in the case $E=0$ where the two-particle  intuition does not work. Now the finiteness of  the negative spectrum of three-particle Hamiltonians $  H$ is determined by so-called zero-energy resonances of two-particle operators $H^\alpha$.
  
  Let us recall this notion; see \cite{LNM}, \S 3.1, for more details. Consider the Schr\"odin\-ger  operator $H=-(2m)^{-1}\Delta + V(x)$ in the space $L^2 ({\mathbb R}^3)$ with a negative non-trivial potential $V(x)  $ which tends to zero faster  than $| x|^{-2}$  as $|x|\to\infty$. To introduce a notion of the zero-energy resonance,
  it is convenient to consider a family of operators 
  \begin{equation}
H(\kappa)= - (2m )^{-1}\Delta +  \kappa V  (x)
  \label{eq:cc}  \end{equation}
  with the coupling constant $\kappa \geq 0$. The operators $ H (\kappa)\geq 0$ for small $\kappa$, and there exists $\kappa_{0}$ such that       for $\kappa> \kappa_{0}$ negative spectrum emerges.
  One says that, for the critical value $\kappa_{0}$, the operator $H   (\kappa_{0})$ has a zero-energy resonance. Its resolvent $  ( H (\kappa_{0})-z I)^{-1}$ has a singularity at the spectral point $z=0$ so that zero-energy resonances are somewhat similar to bound states; sometimes they are even called half-bound states.
  
  It turns out that the discrete spectrum of  three-particle Hamiltonians $ H$ is finite if at most one operator $H^\alpha=-(2m_{\alpha} )^{-1}\Delta +V_{\alpha} (x)$ has  a zero-energy resonance. On the contrary, if two or three  operators $H^\alpha$ have  zero-energy resonances, then the Hamiltonian $   H$  has infinite number of negative eigenvalues. The last result, known as the Efimov effect,  looks counter-intuitive since pair interactions $V_{\alpha}(x_{\alpha})$ between particles may even have finite supports.  
  
Let us give heuristic arguments showing that this surprising result is nevertheless plausible.  For simplicity, we consider identical particles, that is,  $m_{\alpha}=:m$ and $V_{\alpha}=:V$ do not depend on $\alpha$.  For   $\kappa < \kappa_{0}$, operators (\ref{eq:cc}) do not have negative eigenvalues so that the spectrum of the corresponding operator  $H (\kappa)$ coincides with the positive half-axis $[0,\infty)$. 
For   $\kappa > \kappa_{0}$,   negative eigenvalues of the   operators $H^\alpha (\kappa)$ emerge, and hence according to (\ref{eq:HVZ})  the spectrum of the  operator  $ H(\kappa)$ covers the half-axis $[E, \infty)$ where $E= E(\kappa)<0$. It is natural to expect that in the intermediary case $\kappa = \kappa_{0}$, spectral properties of the  operator  $H (\kappa)$  are also intermediary: the essential spectrum is not yet shifted but the infinite discrete spectrum 
appears. 

A precise proof of this fact  given in \cite{Y-Ef} relies on a study of system  (\ref{eq:comp}) for $z\to 0$. Let  $Q(z)$ be  the corresponding matrix operator with non-diagonal components $Q_{\alpha,\beta}(z)$ defined by (\ref{eq:2.11A}) and  $Q_{\alpha,\alpha}(z)=0$. One uses that the strong limit
  $Q(0)$  exists, but  this operator  is not compact. Moreover,  its essential spectrum covers an interval $[0,1+\varepsilon]$ for some $\varepsilon>0$.

\medskip 

  {\it Acknowledgments}. Supported by  RFBR grant No.   20-01-00451A and  Sirius Univ. of Science and Technology (project `Spectral and Functional Inequalities of Math. Phys. and their Appl.')

    \end{document}